\documentclass[fleqn,usenatbib]{mnras}

\usepackage{newtxtext,newtxmath}

\usepackage[T1]{fontenc}

\DeclareRobustCommand{\VAN}[3]{#2}
\let\VANthebibliography\thebibliography
\def\thebibliography{\DeclareRobustCommand{\VAN}[3]{##3}\VANthebibliography}

\usepackage{graphicx}	
\usepackage{amsmath}	
\usepackage{booktabs}

\newcommand{\comment}[1]{}
\newcommand{\lessthan}{\textless \space}
\newcommand{\greaterthan}{\textgreater \space}

\newcommand{\kms}{km~s$^{-1}$}
\newcommand{\kmspace}{km~s$^{-1}$\space}
\newcommand{\kmskpc}{km~s$^{-1}$~kpc$^{-1}$}
\newcommand{\degrees}{$^{\circ}$}
\newcommand{\degreespace}{$^{\circ}$\space}
\defcitealias{Gibson+2023}{G23}

\title[M31 Stellar Chemodynamics]{Towards Understanding the Milky Way's Typicality: Assessing the Chemodynamics of M31's Bulge \& Bar, Thick \& Thin Discs}

\author[B. J. Gibson et al.]{Benjamin J. Gibson,$^{1}$\thanks{E-mail: ben.gibson@utah.edu} 
Gail Zasowski,$^{1}$ 
Anil Seth,$^{1}$ 
Dimitri A. Gadotti,$^{2}$ 
Zixian Wang,$^{1}$ 
\newauthor
Dmitry Bizyaev,$^{3}$ 
Steven R. Majewski,$^{4}$ 
Jon Holtzman,$^{5}$ 
and Sanjib Sharma$^{6}$ 
\\
$^{1}$Department of Physics and Astronomy, University of Utah, Salt Lake City, UT. 84112, USA\\
$^{2}$Centre for Extragalactic Astronomy, Department of Physics, Durham University, South Road, Durham DH1 3LE, UK \\
$^{3}$Apache Point Observatory and New Mexico State University, P.O. Box 59, Sunspot, NM. 88349-0059, USA\\
$^{4}$Department of Astronomy, University of Virginia, Charlottesville, VA. 22904-4325, USA\\
$^{5}$Department of Physics and Astronomy, New Mexico State University, Las Cruces, NM. 88003, USA\\
$^{6}$Space Telescope Science Institute, 3700 San Martin Drive, Baltimore, MD 21218, USA
}

\date{Accepted XXX. Received YYY; in original form ZZZ}

\pubyear{\the\year{}}

\begin{document}
\label{firstpage}
\pagerange{\pageref{firstpage}--\pageref{lastpage}}
\maketitle

\begin{abstract}
We describe a novel framework to model galaxy spectra with two cospatial stellar populations, such as may represent a bulge \& bar or thick \& thin disc, and apply it to APOGEE spectra in the inner $\sim$2~kpc of M31, as well as to stacked spectra representative of the northern and southern parts of M31's disc ($R\sim4-7$~kpc). We use a custom M31 photometric decomposition and A-LIST spectral templates to derive the radial velocity, velocity dispersion, metallicity, and $\alpha$ abundance for both components in each spectrum. In the bulge, one component exhibits little net rotation, high velocity dispersion ($\sim$170~\kms), near-solar metallicity, and high $\alpha$ abundance ([$\alpha$/M]~=~0.28), while the second component shows structured rotation, lower velocity dispersion ($\sim$121~\kms), and slightly higher abundances ([M/H]~=~0.09, [$\alpha$/M]~=~0.3). We tentatively associate the first component with the classical bulge and the second with the bar. In the north disc we identify two distinct components: the first with hotter kinematics, lower metallicity, and higher $\alpha$ abundance than the second ([M/H]~=~0.1 and 0.39, [$\alpha$/M]~=~0.29 and 0.07). These discs appear comparable to the Milky Way's ``thick'' and ``thin'' discs, providing the first evidence that M31’s inner disc has a similar chemodynamical structure. We do not identify two distinct components in the south, potentially due to effects from recent interactions. Such multi-population analysis is crucial to constrain galaxy evolution models that strive to recreate the complex stellar populations found in the Milky Way.

\end{abstract}

\section{Introduction \label{sec:intro}}
Our unique perspective from within the Milky Way (MW) affords us the ability to observe individual stars throughout much of the galaxy. By studying the positions, motions, ages, and chemical makeups of these stars, we can piece together the history of the MW over cosmic time in a way we simply cannot do for external galaxies. This has led to a rise in the number of large scale photometric and spectroscopic surveys of individual stars in the MW. Photometric surveys such as \textit{Gaia} \citep{Gaia_mission}, LSST \citep{LSST}, and DECaPS \citep{DECaPS, DECaPS2} have gathered wide-band, multi-wavelength data for billions of stars throughout the galaxy, providing detailed astrometric, kinematic, and age information for many of them. Surveys like APOGEE \citep{Majewski+2017}, LAMOST \citep{LAMOST}, GALAH \citep{GALAH}, and the SDSS-V Milky Way Mapper \citep[][J.A.~Johnson, in prep]{SDSS-V,Almeida_2023_dr18} have taken spectra of several million stars in the MW and provided detailed chemical abundances and kinematics (collectively known as ``chemodynamics"). 

With these data, we have a detailed understanding of the stellar populations throughout the MW, and we can build models for the formation and evolution of the galaxy. These models are adept at recreating aspects of the galaxy, such as its ``thick" and ``thin" disc dichotomy \citep[e.g.][]{Yoshii1982, Gilmore+Reid1983}, $\alpha$ bimodality \citep[e.g.,][]{Fuhrmann1998, Bensby+2003, Bensby+2005, Reddy+2006, Haywood+2013, Anders+2014, Hayden+2015, Weinberg+2019, Imig+2023}, bulge stellar metallicity distribution \citep[e.g.][]{Zoccali+2008, Rojas-Arriagada+2019, Rojas-Arriagada+2020}, etc. There has been much work to model the formation and evolution of these structures. These models differ by the physical processes they invoke, such as multiple infalls of gas \citep[e.g.][]{Chiappini+1997, Chiappini+2001, Spitoni+2019, Lian+2020a}, stellar radial migration \citep[e.g.][]{Schonrich+Binney2009}, or the dissolution of massive star clumps in the early universe \citep[e.g.][]{Bournaud+2009, Clarke+2019}, to name a few. 

These models are trained primarily on the MW, but may not be applicable to external galaxies, as the MW may not be a typical spiral disc galaxy \citep[e.g.][see Section \ref{sec:litcomp} for more]{Hammer+2007, Licquia+2016, Semenov+2024}. For example, the disc may be unusually compact, which could be the result of the MW having experienced no major mergers in the last $\sim$8-10 Gyr \citep[e.g.][but see also \cite{Lian_2024_MWsize}]{Stewart+2008, Helmi+2018, Myeong+2019, Kruijssen+2019}.

Both \citet{Hammer+2007} and \citet{Licquia+2016} found that M31 was in much better agreement with the Tully-Fisher Relationship and other scaling relations than the MW, so it is reasonable to assume that M31 is a more ``typical" spiral galaxy. Additionally, there is ample evidence that M31 has had a much more active merger history \citep{Ibata+2004, Font+2006, Fardal+2006, Fardal+2008, Sadoun+2014, Dorman+2015, Ferguson+Mackey2016, Hammer+2018, Dsouza+Bell2018}. With this, and given M31's proximity \citep[785 kpc][]{M31Distance} and similarity to the MW, it's an ideal place to investigate the ideas that have influenced our ideas about galaxy evolution for decades.

The introduction of \citet{Gibson+2023}, hereafter referred to as \citetalias{Gibson+2023}, contains a summary of primarily spectroscopic observations of the bulge and inner disc ($\lesssim7$ kpc) of M31. In general, M31 has an old ($>10$ Gyr), solar-metallicity, and $\alpha$-rich (0.2--0.3 dex) bulge containing a more metal-rich bar. Roughly two-thirds of the light in the center comes from a boxy/peanut shaped bulge and bar, while the other one-third comes from a classical bulge \citep{Athanassoula+Beaton2006, Beaton+2007}. The inner disc has super-solar metallicity and near-solar $\alpha$ abundance. 

At larger radii, a number of papers have studied the disc of M31 using individual stellar spectra. For example, \citet{Collins+2011} and \citet{Dorman+2015} investigated how the stellar kinematics correlated with chemistry for their resolved samples, and both found evidence of multiple disc stellar components where the kinematically hotter stars were more metal-poor. 
\citet{Collins+2011} classified stars as belonging to the thick or thin discs by calculating each stars' ``lag" behind a circular velocity field. They found that the thick disc lagged the thin disc by $\sim$46 \kms, that the dispersion and metallicity ([Fe/H]) of the thick disc is $\sim$51 \kmspace and -1.0 dex and these for the thin disc are $\sim$36 \kmspace and -0.7 dex. \citet{Dorman+2015} studied 5800 stars in the northern half of M31 that had Keck-DEIMOS spectra and PHAT photometry \citep{PHAT}. They found that the velocity dispersion and ages of the stars were directly correlated. Additionally, they found that there was a clumpy kinematic component not evenly distributed across the galaxy, and that there is an uniform increase in the velocity dispersion $\sim$6 kpc from the centre that is potentially correlated with the end of the long bar.

JWST NIRSpec observed 103 giant stars in M31 at around 18 kpc from the centre along the disc major axis. These stars were analysed in \citet{Nidever+2024}, who found no indication of a bimodal distribution of $\alpha$ elements in these stars. Instead they found the abundances to vary from [$\alpha$/Fe] of $-0.2$ to almost +0.4 with a near constant number density across the whole sequence, which resembles the abundance pattern of the high-$\alpha$ sequence in the MW, with a larger spread in abundances.

From these studies it is evident that M31 hosts a variety of stellar populations across a range of kinematics and abundances. However, it is unclear just how correlated these components are, in part because the majority of studies of individual stars in M31 focus on the outer disc and halo.

Going beyond the Local Group, several edge-on galaxies have been studied (e.g., with MUSE) to discover how their stellar populations change with distance from the midplane. \citet{Scott+2021} and \citet{Sattler+2023} both found their respective galaxies to have positive vertical $\alpha$ abundance gradients, consistent with the abundance profile of the MW disc. The GECKOS survey \citep{GECKOS} plans to perform this same analysis on 35 more galaxies. As informative as these studies are, they are limited in their ability to measure how the thick and thin discs of these galaxies may overlap at a given point in space, making it difficult to ascertain just how distinct these galaxies' two disc components really are.

So the question remains --- does M31 have multiple spatially-overlapping, chemically- and kinematically-distinct disc components, like its neighbor the MW? In this paper we extend the analysis from \citetalias{Gibson+2023} to characterize the overlapping stellar populations, such as the bulge and bar or thick and thin discs, that exist in M31 by fitting two simple stellar population (SSP) template spectra to each integrated light spectrum. This paper is organized as follows: in Section~\ref{sec:data_model} we describe the spectral data and processing, as well as the models used to analyse it; in Section~\ref{sec:fitting} we describe the spectral fitting methods we used; in Section~\ref{sec:results} we present the results of our analysis and put it in context of the literature; and in Section~\ref{sec:summary} we summarise our results. 

\section{Data and Models}
\label{sec:data_model}

For full details on the observations, data processing, and SSP models used in this paper, see Sections 2.1, 2.2, and 3, respectively, of \citetalias{Gibson+2023}; modifications to the data processing and model interpolation from that paper are highlighted here.

\subsection{Data}
\label{sec:data}
The data for this project were gathered as an ancillary program of the Apache Point Observatory Galactic Evolution Experiment \citep[APOGEE;][S. R. Majewski, in preparation]{Majewski+2017}. APOGEE was part of the third and fourth generations of the Sloan Digital Sky Survey \citep[SDSS-III and -IV;][]{Eisenstein+2011,Blanton+2017}. APOGEE observations were taken by the 2.5-meter Sloan Telescope at Apache Point Observatory \citep{Gunn+2006} and the 2.5-meter du~Pont Telescope at the Las Campanas Observatory \citep{Bowen+Vaughan1973}, both of which are connected to an APOGEE spectrograph \citep{Wilson+2019}. APOGEE data products are processed as described in \citet{Nidever+2015}, and the final data release (DR17) of SDSS-IV is outlined in \citet{Abdurro'uf+2022}. The experiment used $\sim$2\arcsec\space diameter fibers to take high-resolution ($R\sim22,500$), NIR ($1.5 - 1.7 \mu$m) spectra of $\sim$650,000 stars in the MW. APOGEE also observed individual stars in nearby dwarf galaxies and took integrated-light spectra of MW and extragalactic GCs and the bulge and inner disc of M31 \citep{Zasowski+2013, Zasowski+2017, Beaton+2021, Santana+2021}.

APOGEE's wavelength range is in the $H$-band, which contains spectral features from a number of $\alpha$ elements, which trace the chemical evolution and star formation history of a stellar population. Its high spectral resolution enables precise determination of stellar parameters, abundances, and velocity dispersions. Additionally, this allows for efficient subtraction of sky and tellurics. This band is sensitive to RGB and AGB stars and is also subject to low levels of dust attenuation, so APOGEE can see farther into the galactic interior than optical surveys. 

APOGEE observed 1105 locations (henceforth \textit{fiber positions}) in M31 ten times each \citep[each individual observation is called a \textit{visit};][]{Zasowski+2017}. The fiber positions cover roughly 150~kpc$^2$ in the bulge and inner disc of M31, avoiding MW foreground stars and M31 GCs from 2MASS \citep{2MASS} (see Figure~1 from \citetalias{Gibson+2023}).

\begin{figure}
\includegraphics[width=\columnwidth]{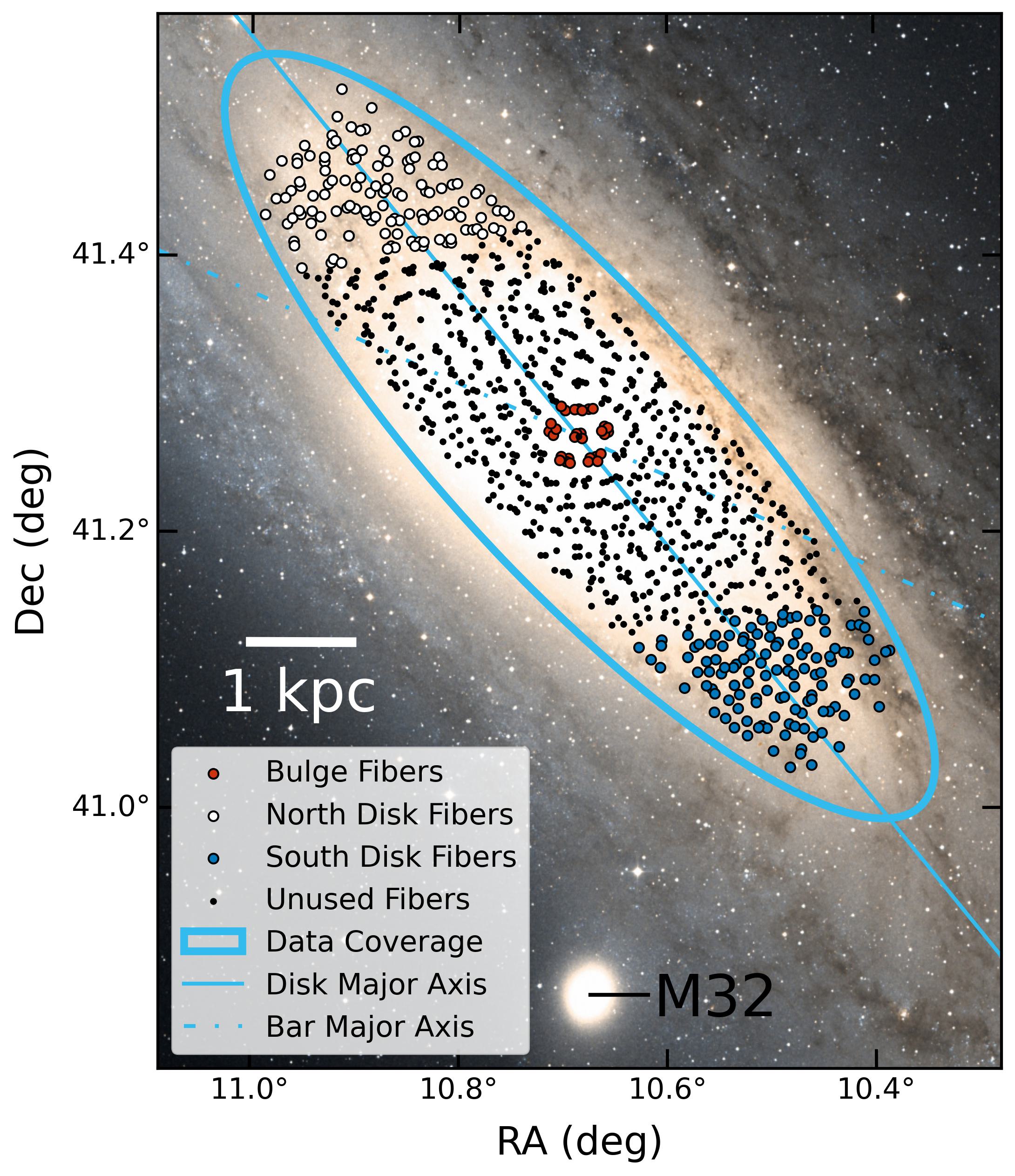}
\caption{APOGEE fiber positions overlaid onto the DSS optical image of M31 (Credit: Infrared Science Archive at IPAC and California Institute of Technology). Bulge fibers (red) were chosen to have a bulge-and-bar-to-total luminosity fraction \greaterthan0.885 and were analysed individually. disc fibers (north: white, south: blue were chosen to have disc-to-total luminosity fractions \greaterthan0.8 and their spectra were co-added before being analysed. The blue ellipse surrounds all the APOGEE fiber positions and is used for reference in future figures. The disc (39.15\degrees) and bar (66.08\degrees) major axes were determined from our custom photometric decomposition described in Section~\ref{sec:decomp}. M32 is shown towards the bottom of the image, and a 1 kpc scale is shown above the legend. \label{fig:fib_pos}}
\end{figure}

Of the 1105 fiber positions, 963 were analysed in \citetalias{Gibson+2023}. For this work, we have narrowed our data sample to 31 fibers in the bulge and 248 in the disc (see Figure~\ref{fig:fib_pos}). The 31 bulge fiber positions have high empirical signal-to-noise ratios (eS/N \textgreater 110, see Section~\ref{sec:prepspec}). The fiber positions in the disc were divided into northern and southern samples, and we co-added their spectra to increase the eS/N to $\sim$90 (see Section~\ref{sec:discspec}). For more detail on how these specific fiber positions were selected, see Section~\ref{sec:decomp}.

\subsubsection{Data Processing from G23 and Modifications}
\label{sec:dataproc}
Our M31 data required additional processing from the standard APOGEE pipeline described in \citet{Nidever+2015}, which is optimized for spectra of individual stars rather than integrated light. For example, the standard data products do not contain low S/N visit spectra, nor spectra with poor radial velocity determinations, both of which were often the case for our M31 observations. The pipeline also did an incomplete job of identifying skylines in our spectra. 

To mitigate these issues, we obtained custom reductions of the M31 observations from the APOGEE data team which did not include the radial velocity determination or the low S/N cutoff. Additionally we created a custom routine for identifying and masking skylines in the spectra. This routine is better suited for use on galactic spectra, which are often fainter than individual stellar spectra and therefore have a worse object flux to sky flux ratio. It also masks pixels affected by other detector effects and works across a wide range of S/N. The full data processing routine is described in detail in \citetalias{Gibson+2023}. Below we present slight modifications to the routine that were made for this work.

In \citetalias{Gibson+2023} we utilized a 500-pixel running median of each spectrum as the psuedo-continuum for normalization. Then we calculated two statistics for each pixel: $\Delta$Flux, which is the absolute value of the flux in the pixel minus the running median, and $S/O$, which is the ratio of the sky flux to the object flux. We identified skylines to be any pixel that is above the 85th percentile in both statistics.

In this work we determined that several individual spectral features in the high-dispersion bulge spectra are broad enough to affect the running median, as a velocity dispersion of 150~\kmspace creates features with a FWHM of 85.3 pixels. Additionally all our features are blended, and so by visual inspection, a typical absorption feature in our spectra is 200 to 250 pixels wide. Therefore, we modified the psuedo-continuum for bulge fiber positions to be a 1000-pixel running median. We still used a 500-pixel running median for disc fiber positions as these spectra have much lower velocity dispersions and S/Ns and therefore their continua are dominated by detector effects such as persistence.

Secondly, we found that the $S/O$ based clipping was very noisy for low S/N spectra, as the sky and object fluxes were often similar in brightness, and so skylines often did not stand out as much as they did in high S/N spectra. Therefore, we now use just the sky flux spectra to identify skylines, and flag pixels that are above the 85th percentile in $\Delta$Flux and above the 60th percentile in sky flux. With this routine we adequately identified and masked pixels in our spectra that have poorly-subtracted skylines.

As in \citetalias{Gibson+2023}, we still masked the ``bump" feature in the green chip and utilized the same selection of APOGEE bitmasks. To combine the psuedo-continuum normalised visit spectra we used an inverse variance weighted average to calculate the flux in each pixel. For the combined fiber spectrum, we masked all pixels that were masked in at least half of the visit spectra.

\subsubsection{Combining disc Fiber Spectra}
\label{sec:discspec}
Individual disc fiber spectra have eS/Ns from one to $\sim$25, so we must co-add fiber spectra to increase their signal. As in \citetalias{Gibson+2023}, we derived a radial velocity measurement for each spectrum using the results from \citet{Opitsch+2018}. We then shifted each to the average radial velocity of the spectra in each bin. In \citetalias{Gibson+2023} we co-added the shifted fiber spectra using the inverse variance weighted average. For this work, we combined the fiber spectra by taking median of the psuedo-continuum normalised fiber spectra and mask pixels that are masked in at least half the visit spectra.

\subsection{SSP Models}
\label{sec:model}

We used the APOGEE Library of Infrared SSP Templates \citep[A-LIST\footnote{\url{https://github.com/aishashok/ALIST-library}};][]{Ashok+2021} to analyse our M31 spectra. A-LIST is a grid of empirical spectral templates made from spectra of MW stars observed by APOGEE. Stellar parameters for the MW spectra were determined using the APOGEE Stellar Parameters and Chemical Abundances Pipeline \citep[ASPCAP;][]{ASPCAP}. A-LIST contains templates spanning a range in age (2-12 Gyr), metallicity (-2.2 $\leq$ [M/H] $\leq$ +0.4) and $\alpha$ abundance (-0.2 $\leq$ [$\alpha$/M] $\leq$ +0.4) and was specifically created for the analysis of integrated-light APOGEE spectra.

\subsubsection{Model Interpolation from G23 and Modifications}
\label{sec:alist_interpolation}

For use in our Markov-Chain Monte Carlo (MCMC) analysis (described in Section~\ref{sec:fitting}), we created a model to interpolate between the discrete A-LIST grid points using \textit{The Cannon} \citep{TheCannon1,TheCannon2}. \textit{The Cannon} is a data-driven machine-learning method to transfer labels (such as independently determined stellar parameters and abundances) from a training set of spectra to a broader data set with unknown labels. This is done by modelling the flux in each pixel as a linear combination of the labels. As such, \textit{The Cannon} can also be used to generate spectra continuously across the label parameter space.

To create our interpolation model, we trained \textit{The Cannon} on a subset of the A-LIST templates. This was largely the same sample as in \citetalias{Gibson+2023} Section~3.2, only in this work we further restricted our training sample and excluded templates with $\alpha$ abundances below 0.0 and above +0.3 dex and $|\Delta T_{\rm eff}| \geq 500$~K \citep[see Section~4.1.2 of][]{Ashok+2021}\footnote{$|\Delta T_{\rm eff}|$ is the difference between the theoretical and measured mean $T_{eff}$ for a given SSP template}. This addressed the occasional fits where MCMC walkers were bifurcating in the parameter space, trapped in local maxima. 

We continued to use the same pixel-weighting scheme as in \citetalias{Gibson+2023} Section~3.2.

\subsection{M31 Photometric Decomposition}
\label{sec:decomp}

Our goal in this paper is to model multiple chemodynamic components in M31's inner regions. We focus this particular study on regions in M31 where we expect two components to dominate the flux, based on the surface brightness profile of the galaxy, and thus perform a structural analysis to characterize the bulge, bar, and disc components.

The structural analysis used an unWISE mosaic at $3.4\mu$m \citep{unWISE1, unWISE2, unWISE3} that included the whole galaxy and enough area around it for a suitable background treatment. The individual images in the mosaic were downloaded directly from \url{https://unwise.me/data/neo7/unwise-coadds/fulldepth/}. To fully cover M31 and its surroundings, we used tiles 0096p393, 0098p408, 0101p424, 0116p393, 0118p408, and 0121p424. The tiles were stitched together using \texttt{reproject.mosaicking.reproject\_and\_coadd()} with \texttt{reproject\_function=reproject\_exact}\footnote{\url{https://reproject.readthedocs.io/en/stable/}}.

A 2D model of flux was fit to the image using IMFIT \citep{IMFIT}, including the unWISE noise image and an oversampled PSF to improve the accuracy of the fits. The Levenberg-Marquardt algorithm was employed to optimize the fit minimizing the $\chi^2$ statistic. All fits included an additional flat background as a free parameter to account for any remaining residual after background subtraction, but this was found to be very small.

The first iteration included an exponential component accounting for the main galaxy disc, and a S\'ersic component to model the boxy/peanut (b/p) bulge. The latter was fit with generalized ellipses to allow it to be boxy. An iterative inspection of the residuals of each of the next fits, which incrementally added more structural components when necessary, indicated that the model required two extra components: a central point source (modeled as a Moffat function), and an additional S\'ersic function also employing generalised ellipses. The latter S\'ersic component accounted for a structure aligned with but beyond each side of the box/peanut bulge, which is presumably the remaining, vertically flat parts of the bar. A new structural component was only kept in the model if it is favoured by the Akaike information criterion.

\begin{figure*}
\includegraphics[width=\textwidth]{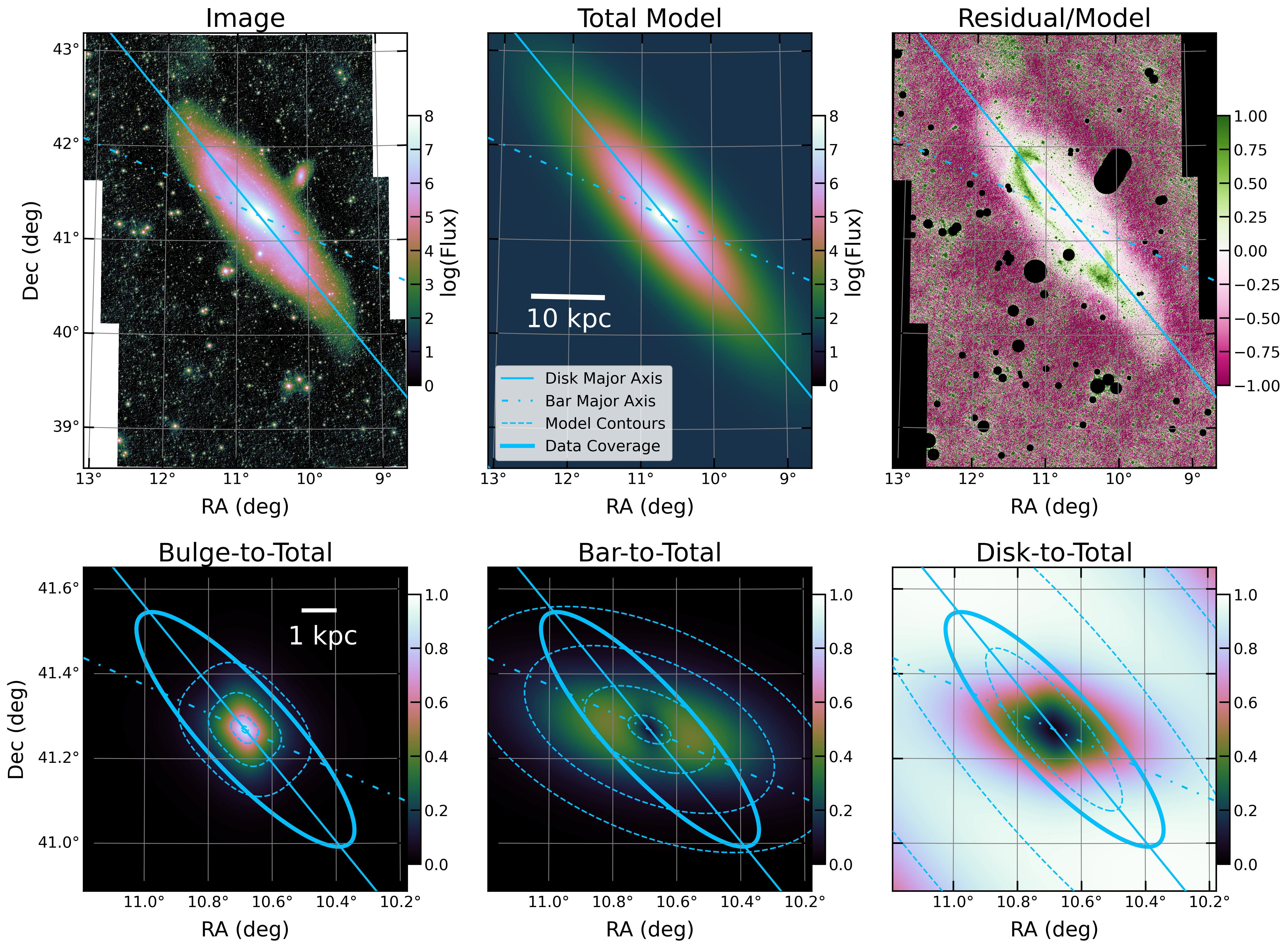}
\caption{Results of the M31 photometric decomposition described in Section~\ref{sec:decomp}. The top left panel shows the unWISE image that was used to perform the decomposition, and the top middle panel shows the total model. The top right panel shows the fractional residuals; black pixels here were masked in the image before performing the fit. The bottom row is zoomed in to the APOGEE data coverage (blue ellipse) and shows the bulge-, bar-, and disc-to-total light ratios with dashed contours indicating the 0.1, 1, 10, and 50\% enclosed light of each individual component. White bars in the top middle and lower left panels indicate the 10 and 1 kpc scales for the top and bottom rows, respectively. \label{fig:decomp}}
\end{figure*}

The final model thus consists of a central point source, the b/p bulge, the disc, and the bar. The latter component is found to be only slightly boxy, but an inspection of the final residuals suggests that part of that component is altered by the possible presence of an inner ring or ansae at the ends of bar (which may explain the low level of boxiness). Values for the parameters of each model component can be found in Table~\ref{tab:decomp_params}, and visualizations of the components can be found in Figure~\ref{fig:decomp}.

\begin{table}
\centering
\resizebox{\columnwidth}{!}{%
\begin{tabular}{@{}l|ccc@{}}
\toprule
Model                            & Parameter   & Value     & Unit           \\ \midrule
Point Source (Moffat Function)   & $\mu_{tot}$ & 12.02   & mag/arcsec$^2$ \\ \midrule
Bulge                            & $\theta$    & 43.6   & degrees        \\
(S\'ersic w/ Generalized Ellipses) & $e$         & 0.26    & -              \\
                                 & $c_0$       & 0.71    & -              \\
                                 & $n$         & 1.59    & -              \\
                                 & $\mu_e$     & 14.98   & mag/arcsec$^2$ \\
                                 & $r_e$       & 131.5  & arcsec         \\ \midrule
Bar                              & $\theta$    & 66.1   & degrees        \\
(S\'ersic w/ Generalized Ellipses) & $e$         & 0.48    & -              \\
                                 & $c_0$       & 0.03    & -              \\
                                 & $n$         & 0.95    & -              \\
                                 & $\mu_e$     & 16.98   & mag/arcsec$^2$ \\
                                 & $r_e$       & 405.8  & arcsec         \\ \midrule
Disc                             & $\theta$    & 39.1   & degrees        \\
(Exponential)                    & $e$         & 0.72    & -              \\
                                 & $\mu_0$     & 16.27   & mag/arcsec$^2$ \\
                                 & $h$         & 1229.3 & arcsec         \\ \bottomrule
\end{tabular}}
\caption{Table of photometric decomposition model parameters. Surface brightnesses $\mu$ were converted from counts per pixel to mag/arcsec$^2$ using a 3.6$\mu$m reference surface brightness of 11.5 mag/arcsec$^2$ at the centre of M31 from \citet{Barmby+2006}. $\theta$ is the position angle of the model component measured in degrees east of north; we use these values for the disc and bar for plots and gradient calculations throughout this paper. $e$ is the ellipticity of the model component. $c_0$ governs the boxiness of the function, $n$ is the S\'ersic index, $r_e$ is the effective radius, and $h$ is the scale length of the disk. \label{tab:decomp_params}}
\end{table}

We used this decomposition in two ways. The first was to select which fiber positions to analyse. In the bulge we chose all fiber positions where the bulge plus the bar luminosity accounts for \greaterthan88.5\% of the total light. We note that there is at least a minor contribution from a classical bulge component in this region of the galaxy in addition to the b/p bulge and bar, though this has not been captured in this decomposition. We excluded the fiber position at the very centre, as it has \textgreater50\% contamination from the central point source. Our final bulge sample consists of 31 fiber positions within 91.5" of the centre of M31. Fiber positions beyond this radius have a significant disc contribution and we want to isolate regions where there are only two dominant components. 

In the disc we selected fibers that have a disc-to-total ratio of \greaterthan80\% and split them into two samples. The northern-disc sample contains 125 fiber positions and the southern-disc sample contains 123; both samples' spectra were co-added to produce a single high-S/N spectrum. The fiber positions in all three samples are shown in Figure~\ref{fig:fib_pos}. We fit all of our spectra with two chemodynamic components and test whether these fits provide significantly better matches to the data than one chemodynamic component as described in Section~\ref{sec:fitting}. 

Secondly, we used this decomposition as a prior on the light weighting fraction $f$ for bulge fiber positions, which is a fitting parameter described in Section~\ref{sec:twocomp}. Specifically, we calculated $f$ to be the ratio of bulge luminosity to the bar plus the disc luminosity, under the assumption that the bar and disc have, on average, more similar stellar chemodynamics than either do to the bulge:
\begin{equation}
    f = 1 - \frac{L_{bar} + L_{disc}}{L_{bulge}}. \label{eq:frac}
\end{equation}

\section{Spectral Fitting}
\label{sec:fitting}

To analyse our spectra we utilized a Markov-Chain Monte Carlo (MCMC) full-spectrum fitting routine to identify the best fit to our data. We employed the \texttt{emcee} software package \citep{emcee} to run the MCMC and determine the set of parameters that has the maximum likelihood, $L = -\chi^2/2$. At each iteration, one or two template spectra were generated from the interpolated A-LIST model (Section~\ref{sec:alist_interpolation}) and shifted and broadened using \texttt{ppxf.ppxf\_util.convolve\_gauss\_hermite()}. Finally, the template continuum was calculated by taking a 1000- or 500-pixel running median (for the bulge and disc, respectively) before evaluating the $\chi^2$ of the fit.

\subsection{Preparing Spectra for Fitting}
\label{sec:prepspec}
Before fitting each spectrum, we masked out pixels with untrustworthy fluxes, properly scaled the error arrays, and corrected for the continuum.

As in \citetalias{Gibson+2023} (Section~4.2), we used the masks generated in the data processing routine. We also masked the 250 pixels closest to the red end of each chip and the 100 pixels closest to the blue end of each chip. Lastly, for the disc spectra we masked pixels between 15360-15395, 16030-16070, and 16210-16240 \AA, as these regions were found to be poorly fit by our models (see Section~5.5.5 in \citetalias{Gibson+2023}).

Next, we calculated the empirical signal-to-noise ratio (eS/N) by using \textsc{ppxf} \citep[Penalized PiXel-Fitting,][]{pPXF} to fit each chip individually with \texttt{mdegree} set to 40; this parameter is the order of the multiplicative polynomial that \textsc{ppxf} uses to model the continuum of a spectrum. We used a high \texttt{mdegree} to produce a ``perfect" fit to the data with no structure in the residuals. We loaded templates generated at all the discrete A-LIST grid points that were used to generate the model (see Section~\ref{sec:alist_interpolation}) into \textsc{ppxf}. From this fit, we calculate the normalized interquartile range, $\sigma_G$, of the residuals divided by the uncertainties ($q$):
\begin{equation}
 \sigma_G = 0.7413 ( q_{75} - q_{25} )
\end{equation}
where $q_{75}$ and $q_{25}$ are the upper and lower quartiles of $q$, and the 0.7413 term is used to scale the interquartile range with that of a Gaussian distribution. We multiplied the uncertainty arrays by $\sigma_G$ to make the $\chi^2$ value for this fit equal one.

Lastly, we corrected for the pseudo-continuum of each spectrum. In the bulge we did this by subtracting the median residual of a number of high-eS/N, one-component fits from the spectrum; in the disc we subtract the median residual between a one-component fit to the combined spectrum and each individual fiber spectra. Both methods are described more fully in Section~\ref{sec:continuum}.

\subsection{One-Component Spectral Fits}
\label{sec:onecomp}

We first fit each fiber spectrum within the bulge region with a single template spectrum to identify the single best fitting set of parameters, which were taken to be light-weighted average chemodynamics of the stellar population at each fiber position. Our model parameters are radial velocity ($V$), velocity dispersion ($\sigma$), metallicity ([M/H]), $\alpha$ abundance ([$\alpha$/M]), and age. We fixed the age to 10 Gyr, as a) our SSP models are not very sensitive to age variations at $\gtrsim$ 8 Gyr (as shown in \citetalias{Gibson+2023}), and b) the average age of the stellar populations dominating the NIR in the bulge has been shown to be at least 10 Gyr \citep[e.g.,][]{Olsen+2006, Saglia+2010, Saglia+2018}. 

We measured the other parameters using the MCMC routine with 100 walkers for 300 iterations and \texttt{StretchMove} with \texttt{a}=8.0\footnote{\texttt{StretchMove} is the default emcee ``move": an algorithm that governs how walkers can move throughout the parameter space. The \texttt{a} parameter sets the average distance a walker can move from iteration to iteration. The default is 2.0}. The ranges between which we initialized the MCMC walkers and the bounds for the parameter space can be found in Table~\ref{tab:MCMC_bounds}. Ranges for the kinematics are listed in the table itself, with ranges for the abundances listed in the caption. After removing a burn-in period of the first 150 iterations, we quote the median values of the ensemble chain as the final, best-fitting result for each parameter. Any time we refer to the results of a one-component fit, we call this the ``mean" value for the parameter, so for example the one-component metallicity of a given spectrum would be the mean metallicity, $\rm [M/H]_{mean}$, of that spectrum's stellar population.

\begin{table*}
\centering
\resizebox{\textwidth}{!}{%
\begin{tabular}{@{}cc|cc|cccc@{}}
\toprule
                      &                 & \multicolumn{2}{c|}{One-Component Fits} & \multicolumn{4}{c}{Two-Component Fits}          \\ \midrule
                      &                 & $V$                  & $\sigma$         & $V_1$     & $V_2$     & $\sigma_1$ & $\sigma_2$ \\ \midrule
\multicolumn{1}{c|}{Bulge}         & Initialization  & (-350, -250) & (60, 190) & (-400, -200) & (-400, -200) & (150, 200) & (100, 150) \\
\multicolumn{1}{c|}{} & Parameter Space & (-400, -200)         & (10, 220)        & (-600, 0) & (-600, 0) & (10, 220)  & (10, 220)  \\ \midrule
\multicolumn{1}{c|}{Northern disc} & Initialization  & (-150, -100) & (50, 150) & (-300, 0)    & (-300, 0)    & $\sigma_{\rm mean}\pm10$   & $\sigma_{\rm mean}\pm10$   \\
\multicolumn{1}{c|}{} & Parameter Space & (-300, 0)            & (10, 220)        & (-300, 0) & (-300, 0) & (10, 220)  & (10, 220)  \\ \midrule
\multicolumn{1}{c|}{Southern disc} & Initialization  & (-500, -450) & (50, 150) & (-600, -300) & (-600, -300) & $\sigma_{\rm mean}\pm10$   & $\sigma_{\rm mean}\pm10$   \\
\multicolumn{1}{c|}{}              & Parameter Space & (-600, -300) & (10, 220) & (-600, -300) & (-600, -300) & (10, 220)  & (10, 220)  \\ \bottomrule
\end{tabular}%
}
\caption{Bounds for the initialization and total explorable parameter space for kinematic components in the MCMC. All values are given in \kms. Values with the subscript $_{\rm mean}$ are the one-component results for that parameter.\\
All one-component [M/H] and [$\alpha$/M] are initialized between (-0.8, +0.3) and (+0.05, +0.25), respectively. All two-component [M/H] and [$\alpha$/M] are initialized within $\pm0.1$ dex of $\rm [M/H]_{mean}$ and $\rm [\alpha/M]_{mean}$. The bounds of the [M/H] and and [$\alpha$/M] parameter spaces for both one- and two-component fits are always (-0.8, +0.4) and (+0.05 + +0.4), respectively.\label{tab:MCMC_bounds}}
\end{table*}

\subsection{Two-Component Spectral Fits}
\label{sec:twocomp}

Next, we fit each spectrum with a linear combination of two template spectra, $T_{\rm comb}$, weighted by the fractional light contribution, $f$ (see Equation~\ref{eq:frac}, from each component:

\begin{equation}
    T_{\rm comb} = f T_1 + (1-f) T_2.
\end{equation}

This allows us to individually characterize the two dominant stellar populations that make up the integrated light at our fiber positions. With this combination, we almost doubled the number of parameters in our MCMC: separate velocities, dispersions, metallicities, and $\alpha$ abundances for our two models. We continued to fix the age of each model to 10~Gyr.

For the two-component fits, we invoked multiple priors on the kinematics. Primarily, we require Component 1 to be slower-rotating and more dispersion-dominated than the other. This is a natural consequence of two populations that exist in the same potential and has been shown to exist between the MW thick and thin discs \citep[e.g.,][]{Mackereth_2019_diskheating,Vieira_2022_MWdisks} and the bar and bulge \citep[e.g.,][]{Ness_2016_apogeekinematics,Zasowski+2016,Queiroz_2021_MWbar}. Specifically, we require that Component 1 has a slower rotation velocity ($V_1$) than both the mean ($V_{\rm mean}$) and Component 2 ($V_2$). Practically, this is done using Equation~\ref{eq:vels}, where the systemic velocity of M31 is $-300$~\kmspace \citep{M31velocity}. We further require that the velocity of each component be related to the mean velocity by the fractional light-weighting of each component ($f$) squared by Equation~\ref{eq:frac_vel}. Lastly, we require the dispersion of Component 1 ($\sigma_1$) to be higher than the dispersion of Component 2 ($\sigma_2$). These three priors are thus defined by:
\begin{equation}
    |V_1 + 300| < |V_2 + 300| \textrm{ and } |V_{\rm mean} + 300| \label{eq:vels}
\end{equation}
\begin{equation}
    f^2 V_1 + (1-f^2) V_2 = V_{\rm mean} \pm 12 \label{eq:frac_vel}
\end{equation}
\begin{equation}
    \sigma_1 > \sigma_2
\end{equation}
Equation~\ref{eq:frac_vel} was derived empirically using fits to simulated observations. The $\pm12$ in this equation was chosen to encapsulate the variation in results from the derivation (see Section~\ref{sec:mock_obs} for details). We emphasize that while the kinematics of the two components are coupled, the metallicity and $\alpha$ abundances are allowed to vary completely independently. 

\subsubsection{Fitting Bulge Spectra}
\label{sec:bulgefits}

For the two-component MCMC routine, we analyse 100 walkers for 700 iterations with the same \texttt{StretchMove} algorithm as in the single-component fits (Section~\ref{sec:onecomp}). Again, bounds for the kinematic components are shown in Table~\ref{tab:MCMC_bounds}, with the abundance bounds in the caption. [M/H] and [$\alpha$/M] are initialized within $\pm0.1$ of the mean results. After removing a burn-in period of the first 300 iterations, we again quote the result for each parameter to be the median value of the ensemble chain. We then perform a jackknife resampling routine \citep{jackknife} of the visit spectra in order to determine the observational errors on our measurements, as described in \citetalias{Gibson+2023} Section~4.3. Results from these two-component fits are discussed in Section~\ref{sec:bulge_results} below.

\subsubsection{Fitting Combined disc Spectra}
\label{sec:discfits}

For the two combined disc spectra (``north'' and ``south''; Figure~\ref{fig:fib_pos}), we used the same overall fitting routine as for the bulge spectra, with the same set of two-component kinematic priors (Section~\ref{sec:bulgefits}). Given the rotation of M31, we changed the walker initialization and the bounds of the physically plausible parameter space for the kinematics, as these are different on either side of the disc. The kinematics bounds are shown in Table~\ref{tab:MCMC_bounds}. We use the same bounds for the metallicity and $\alpha$ abundance as before.

The primary change to our analysis of the disc spectra was that rather than fixing the light weighting fraction $f$ to a value determined from the photometric decomposition, we systematically varied $f$ in increments of 0.1 between 0.2 and 0.8 and analysed the data separately with each value. This was done so that we can quantify the unknown weighting of the two disc components.

For both sides of the disc and at each value of $f$, we ran the MCMC with 100 walkers for 700 iterations and removed a burn-in period of the first 300 iterations. We again quoted the result for each parameter as the median value of the ensemble chain and jackknifed the fiber spectra to determine the errors on our measurements.

\subsection{Adjusting Continuum}
\label{sec:continuum}

In \citetalias{Gibson+2023} we handled the continuum prior to visit-stacking by normalizing each visit spectrum with a 500-pixel running median and then applying a 2nd order multiplicative polynomial to each template using \textsc{ppxf} by setting \texttt{mdegree}~$=2$. As stated in Section~\ref{sec:dataproc}, in this work we instead normalized each visit before combination as well as each template during the MCMC with a 1000- or 500-pixel ($\sim$220 and $\sim$110 \AA) running median for the bulge and disc, respectively. This change was prompted by the realization that a 500-pixel running median is not wide enough to be unaffected by the broadest absorption features in our spectra, which are 200-250 pixels wide. Additionally, we wanted to more finely control the normalization applied to each template spectrum than we could with the pPXF-determined polynomial used previously.

Even after these modifications, the observed spectra present large continuum discontinuities that are not reproduced in the SSP models. In \citetalias{Gibson+2023}, we addressed these discontinuities with a low-order polynomial applied to the model spectra during fitting (see Section 4.2 of that paper). However, because of the improved MCMC fitting routine here, which applies the same continuum to each template, we cannot address the discontinuities during fitting, and in effect the $\chi2$ of our fits are dominated by these regions, rather than by any improvements made by fitting with two template spectra. Therefore, in this work we empirically modify the continua of our spectra as outlined below.

In the bulge we took the 81 fiber positions where the bulge-plus-bar luminosity accounts for $>$80\% of the total light and fit their spectra with the one-component fitting routine outlined in Section~\ref{sec:onecomp}. We took the residuals from each of these fits and calculate the median value for each pixel, ignoring masked pixels. We then subtract these median residuals from the flux spectrum of each fiber position before fitting again. We tested both Doppler shifting the residuals to the rest wavelength before taking the median, and multiplying the spectra by the median residuals (shifted to centre on unity). We found that the observed-frame residuals gave velocity results more consistent with the unmodified spectra, and that subtracting the residuals gave lower $\chi^2$ values than multiplying by them. 

\begin{figure}
\includegraphics[width=\columnwidth]{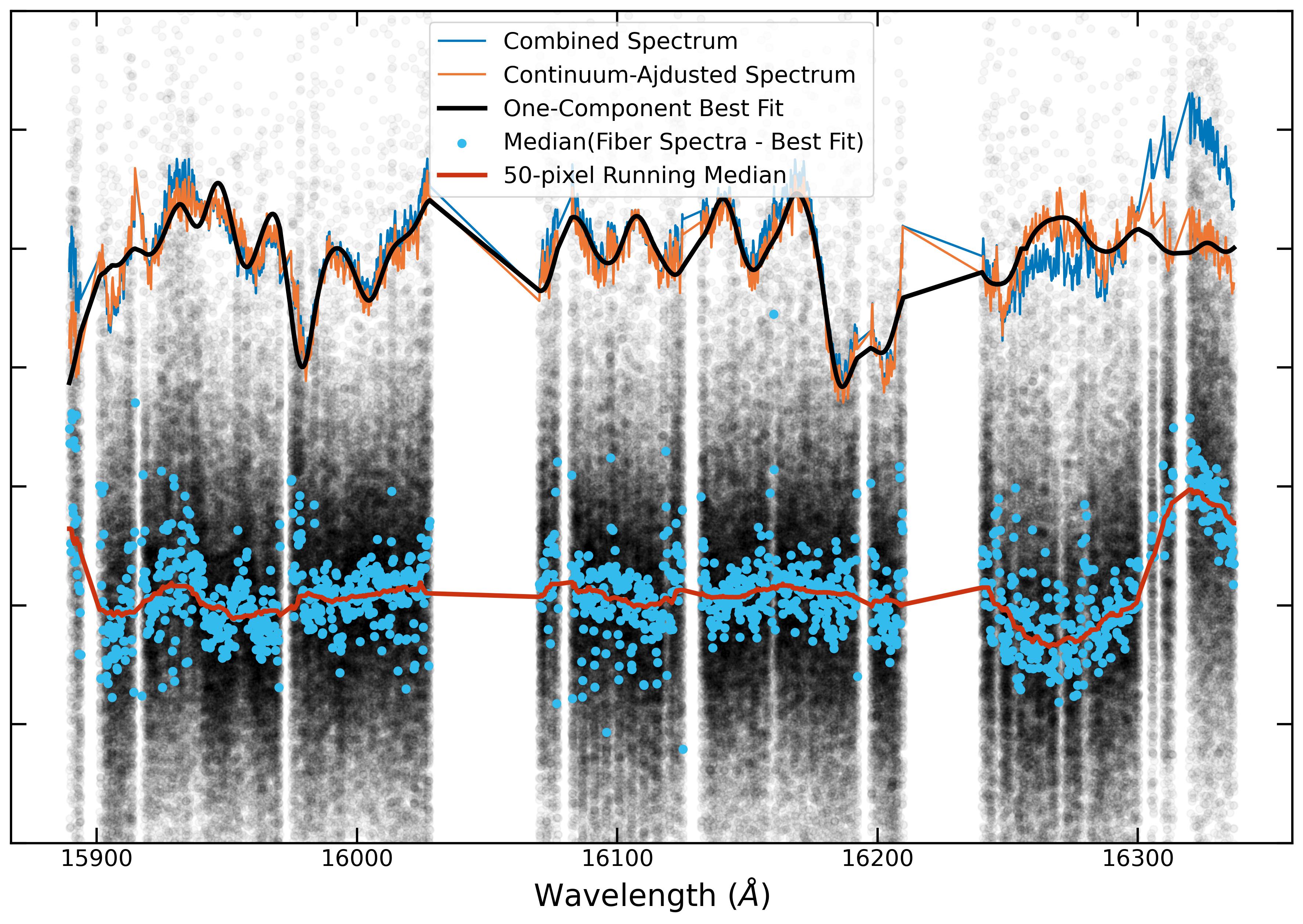}
\caption{Demonstration of the continuum adjustment in the green chip for the northern combined disc spectrum. The dark blue and orange spectra are the unadjusted and adjusted combined spectrum, respectively. The one component fit to the unadjusted spectrum is shown as the black line. Gray translucent points show the residuals between the individual fiber spectra in the north sample and the one component fit to the combined north disc spectrum. The light blue points are the median of those residuals, and the red line is the 50-pixel wide running median that was used to adjust the continuum.\label{fig:disc_cont}}
\end{figure}

In the disc, we performed a one-component fit to each of the combined spectra, then calculated the residuals between the best fit and each of the fiber spectra used to create the combined spectrum as shown in Figure~\ref{fig:disc_cont}. As with the bulge spectra, we calculated the median residual in each pixel, again ignoring masks. This median residual had general patterns but was noisy, given that the individual fiber spectra often have eS/Ns below five. Therefore we took a 50-pixel running median of the median residual to smooth over the noise and applied that to the combined spectra. Once again we used the observed frame residuals and subtracted them from the combined spectrum.

\subsection{Validation of Methodology}

\subsubsection{Mock Observations}
\label{sec:mock_obs}
To validate our methodology, we tested it on mock observations of M31: composite template spectra with realistic noise added. To create these mock observations, we generated two template spectra with different metallicities, $\alpha$ abundances, and age of 10~Gyr, then shifted and broadened them by different, but realistic, velocities and dispersions using \texttt{ppxf.ppxf\_util.convolve\_gauss\_hermite()}. We then added the two templates together, weighting them by some fraction $f$. We then took a random high-eS/N spectrum ($\gtrsim80$) and interpolated it onto the same wavelength array as the combined template. We took the average sky flux from all the visit spectra for the fiber position and added it to the combined templates, added Gaussian noise corresponding to the uncertainty array, then subtracted the sky flux back out before applying the skyline masking routine from Section~\ref{sec:dataproc}. The end result are realistic APOGEE mock observations of two co-spatial stellar populations.

We used these mock observations for two separate tests. The first test was used to derive the fractional velocity relationship (Equation~\ref{eq:frac_vel}), and the second was to verify that our fitting methods return the known input parameters.

\begin{figure}
\includegraphics[width=\columnwidth]{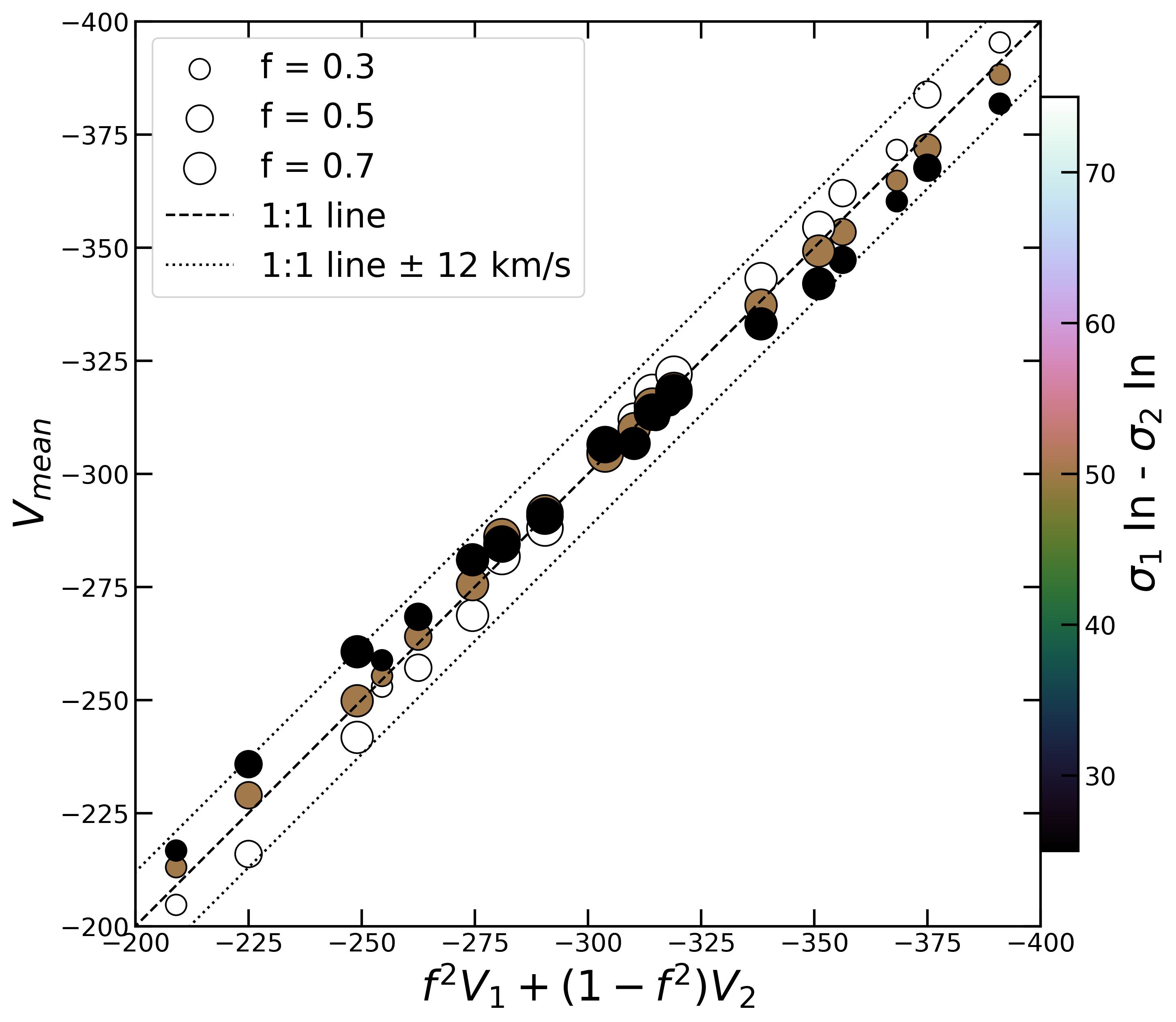}
\caption{Validation of Equation~\ref{eq:frac_vel}. 60 mock observations with eS/N $\simeq100$ were generated across a range of velocities, dispersions, and fractions and were fit with the same routine outlined in Section~\ref{sec:fitting}. Points are coloured by the difference in the input dispersions for the two components and sized by the light-weighting fraction. There is some correction that could be applied given the input dispersions, but the behavior is constrained by allowing the calculated velocity to exist in a range $\pm$12 the mean value. \label{fig:frac_vel}}
\end{figure}

For the first test we generated two-component mock observations with eS/N $\simeq100$ (similar to the bulge fiber spectra) across a wide range of realistic velocities and dispersions for each component, keeping the abundances different for each component but consistent across all mock observations. We then fit these mock observations with the one-component fitting routine outlined in Section~\ref{sec:onecomp}. We found that the mean velocity is consistent with the linear combination of the two velocities weighted by their light fraction, $f$ (Equation~\ref{eq:frac_vel}). This behavior is shown in Figure~\ref{fig:frac_vel}.

\begin{figure}
\includegraphics[width=\columnwidth]{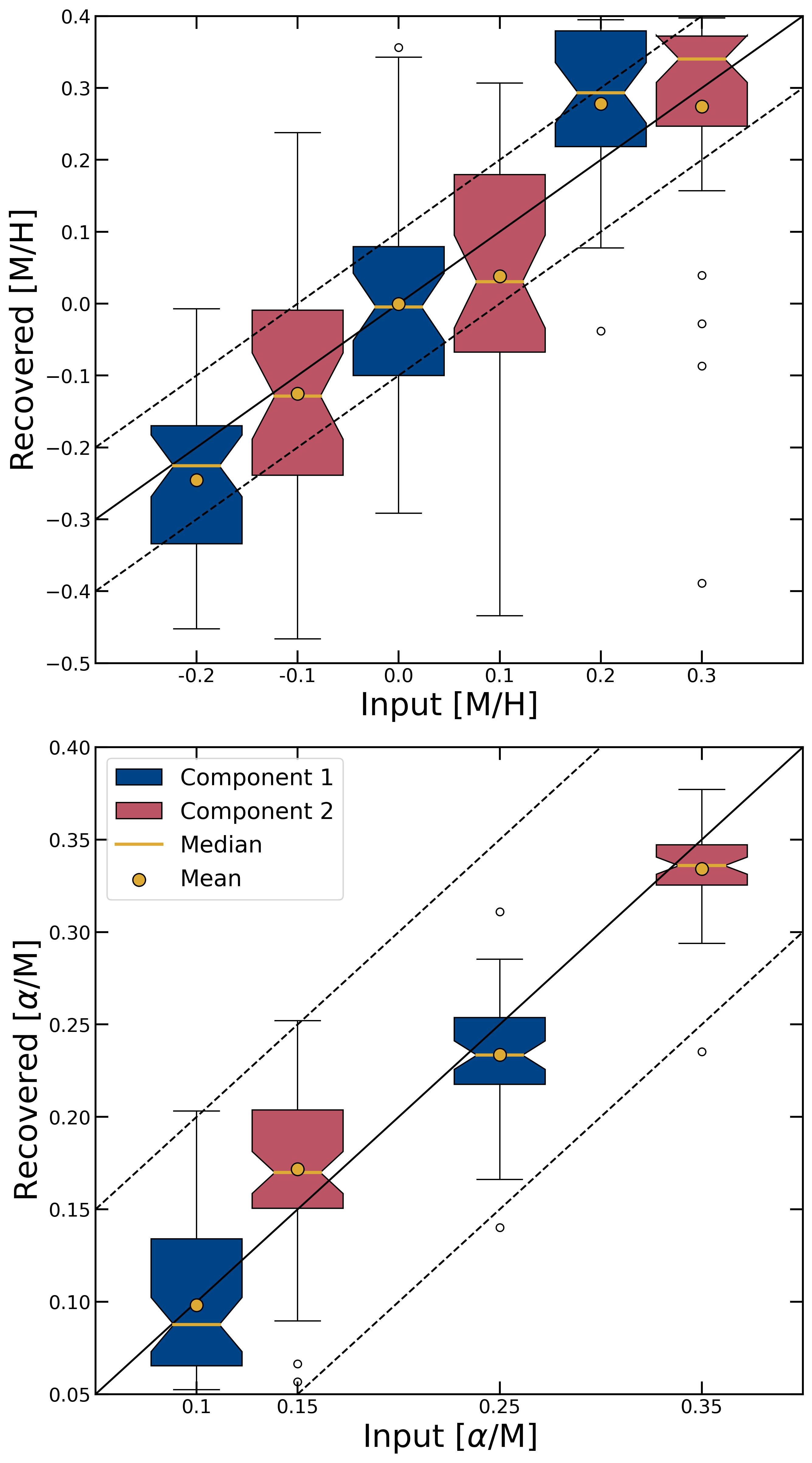}
\caption{Validation using mock observations that the fitting routine outlined in Section~\ref{sec:fitting} is able to recover input abundances to within $\sim$0.1 dex (dashed diagonal lines).\label{fig:recovery}}
\end{figure}

Once this relationship was known, we then performed the second test, integrating Equation~\ref{eq:frac_vel} into our methods. We generated another set of two-component mock observations with varying $f$. This time all mock observations had the same realistic kinematics but numerous combinations of metallicity and $\alpha$ abundance. We then ran the full fitting routine from Section~\ref{sec:fitting}, varying $f$ as we did in Section~\ref{sec:discfits} on the mock observations, and found we were able to recover the input fraction 63\% of the time. When fixing $f$ to the input value, we recovered input abundances within roughly 0.1~dex in almost all tests. These results are shown in Figure~\ref{fig:recovery}.

\subsubsection{How well can we recover light fractions?}
\label{subsec:fraction}

For our fiber positions in the bulge, we set the light weighting fraction $f$ to a value determined by the photometric decomposition of Section~\ref{sec:decomp}. However, given that our decomposition matches the bulk properties of M31, it's possible that the fraction determined at a fiber position is not accurate given that a small number of stars actually make up the light there. So we tested how large this shot noise could be by generating two synthetic stellar populations across a range of metallicities, $\alpha$ abundances, fractions, and masses.

At each combination of inputs we generated two isochrones using \texttt{SPISEA} \citep{SPISEA} and populated each with stars. We then used a magnitude cut to isolate the giants (since our APOGEE data is dominated by such stars). The number of stars in each population ranged from a few hundred to a few thousand depending mostly on the mass of the populations. We then combined the two populations, selected 100 stars at random, and calculated the number that originated from each population. We found that the number of stars selected was within 10\% of the input fraction 85 to 95\% of the time across all combinations of input parameters. In other words, the actual fraction of stars coming from each population in our M31 bulge fiber positions should be within 10\% of the fraction determined by our photometric decomposition up to 95\% of the time.

\subsubsection{Two- vs. One-Component Fit Quality}
\label{sec:deltabic}

\begin{figure}
\includegraphics[width=\columnwidth]{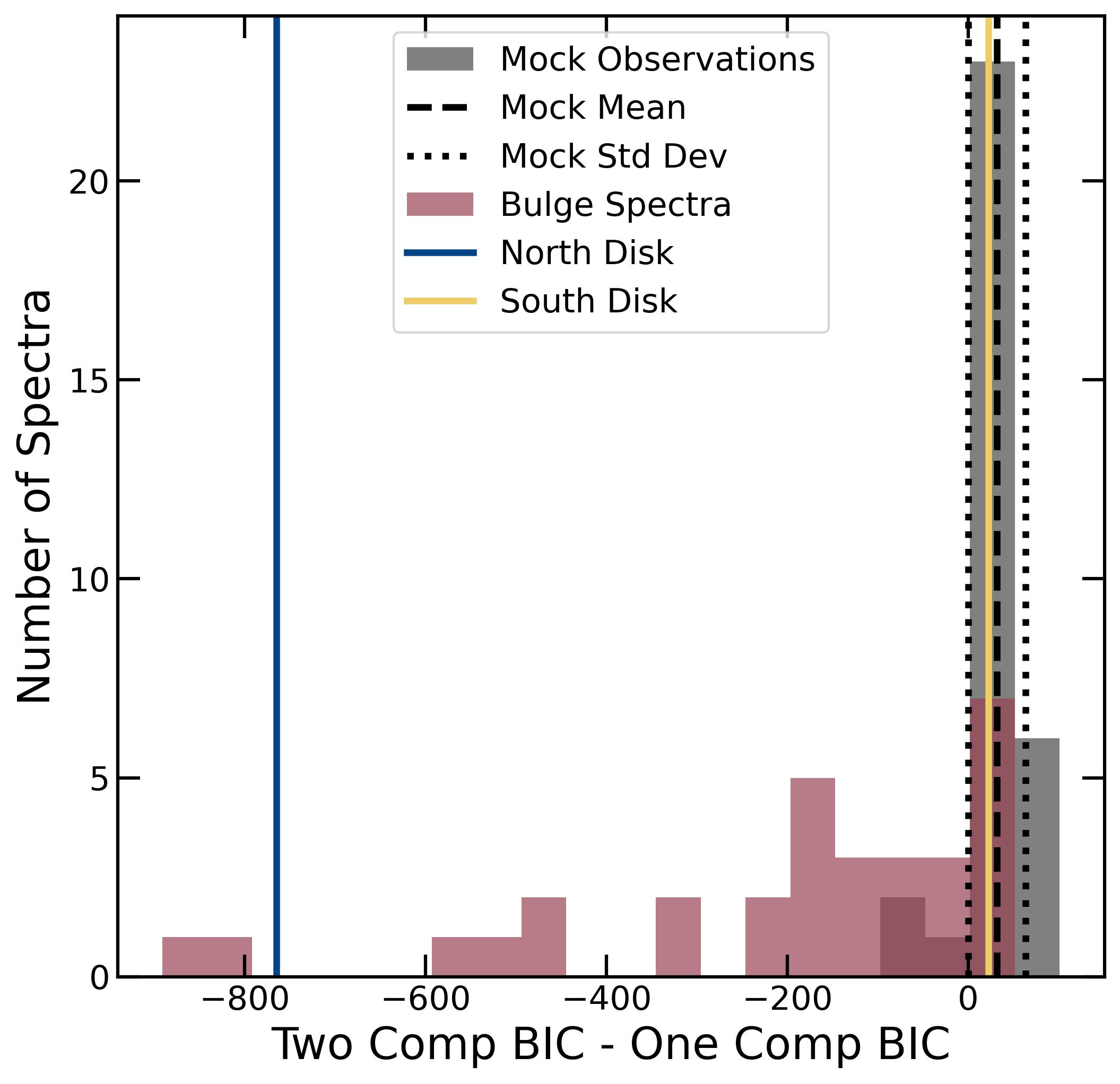}
\caption{Comparison of the BIC between two-component fits to the 31 bulge spectra (red) and 32 one-component mock observations (gray). The mean and standard deviation $\Delta$BIC for the mock observations are shown as black dashed and dotted lines, respectively. The $\Delta$BIC for the combined disc spectra are also shown as solid vertical lines. In general, we recover that the mock observations only have one component. We also find that many of the two-component fits to real data are statistically better than one-component fits to real data.\label{fig:delta_bic}}
\end{figure}

To test whether our one- or two-component fits better model M31's bulge and disc spectra, we calculated the Bayesian Information Criterion (BIC) of both fits. The difference between these values ($\Delta$BIC) are shown in the teal histogram in Figure~\ref{fig:delta_bic}. The negative values found for most spectra indicate that the two-component fits are a significant improvement over the one-component fits to real data, despite having twice the number of parameters. 

As a comparison point to the real data, we generated a grid of one-component mock observations and fit them with one and two components, again comparing the BICs. These are shown as the gray histogram in Figure~\ref{fig:delta_bic}, and demonstrate the range of $\Delta$BIC values we might expect for real spectra that intrinsically only have one component. The mean value of the BIC here is slightly positive (31.71 with a standard deviation of 31.77, black dashed and dotted lines), which is expected given the penalty paid for extra parameters in the BIC. The spread on these BIC values gives us an indication of how significant our derived BICs for the real data are; in 22 of the 31 bulge fibers and the northern combined disc spectrum, we find that the $\Delta$BIC is more than 2$\sigma$ below the BIC for the one-component mock observations, suggesting we are in fact characterizing two distinct chemodynamic components in our M31 data. The southern combined disc spectrum ($\Delta\textrm{BIC}=22.29$) does not satisfy the 2$\sigma$ $\Delta$BIC cutoff, indicating that we are unable to recover two chemodynamic components from our data. See Section \ref{sec:discresults} for more details.

\section{Results and Discussion}
\label{sec:results}

Table~\ref{tab:results} shows the rotational velocity, dispersion, metallicity, and $\alpha$ abundance for each component in the three regions of M31 we investigated, each of which is described more fully below.

\begin{table*}
\resizebox{\textwidth}{!}{%
\begin{tabular}{@{}ll|cccccl@{}}
\toprule
Region     & Component   & $f$         & $V_{rot}$          & $\sigma$           & {[}M/H{]}       & \multicolumn{1}{c|}{{[}$\alpha$/M{]}} & Associated Substructure    \\ \midrule
           & Mean        & -           & $28.00\pm16.22$  & $162.58\pm12.19$ & $0.04\pm0.04$ & \multicolumn{1}{c|}{$0.29\pm0.02$}  & -                        \\
Bulge      & Component 1 & $\sim$0.68 & $27.683\pm19.59$  & $169.42\pm15.67$ & $0.02\pm0.09$ & \multicolumn{1}{c|}{$0.28\pm0.05$}  & Classical Bulge          \\
           & Component 2 & $\sim$0.33 & $84.73\pm33.162$  & $121.76\pm26.47$ & $0.09\pm0.12$ & \multicolumn{1}{c|}{$0.30\pm0.05$}  & Bar          \\ \midrule
           & Mean        & -           & $179.71\pm4.24$  & $80.94\pm4.52$   & $0.20\pm0.07$ & \multicolumn{1}{c|}{$0.24\pm0.02$}  & -                        \\
North disc & Component 1 & 0.8         & $157.64\pm9.49$  & $74.03\pm7.31$   & $0.10\pm0.08$ & \multicolumn{1}{c|}{$0.29\pm0.03$}  & $\alpha$-rich Thick disc \\
           & Component 2 & 0.2         & $246.86\pm10.90$ & $41.19\pm18.72$  & $0.39\pm0.04$ & \multicolumn{1}{c|}{$0.08\pm0.13$}  & $\alpha$-poor Thin disc  \\ \midrule
South disc & Mean        & -           & $181.95\pm5.27$  & $88.52\pm4.43$   & $0.09\pm0.05$ & \multicolumn{1}{c|} {$0.29\pm0.02$}                       & -                        \\ \bottomrule
\end{tabular}%
}
\caption{Chemodynamics results for the three regions analysed. In the bulge, results are averaged over all 22 fibers that meet the $\Delta$BIC cutoff as defined in Section~\ref{sec:deltabic}. The galactic substructure that is potentially associated with each component is listed in the right-most column.\label{tab:results}}
\end{table*}

\subsection{The Bulge Region}
\label{sec:bulge_results}

The bulge region contains four fiber positions with a galactocentric radius within 0.04 kpc and 27 fiber positions in an annulus from 0.235 to 0.350 kpc. For the following results, we removed the nine fiber positions that did not meet the $\Delta$BIC cutoff as defined in Section~\ref{sec:deltabic}. These fiber positions are those that do not have thick black outlines in Figure~\ref{fig:bulge_kins}. All gradients presented have been calculated with these fiber positions excluded. We note that the removal of these positions barely change the numbers we quote for these gradients, and the qualitative interpretations of them are the same.

\subsubsection{Kinematics}

\begin{figure}
\includegraphics[width=\columnwidth]{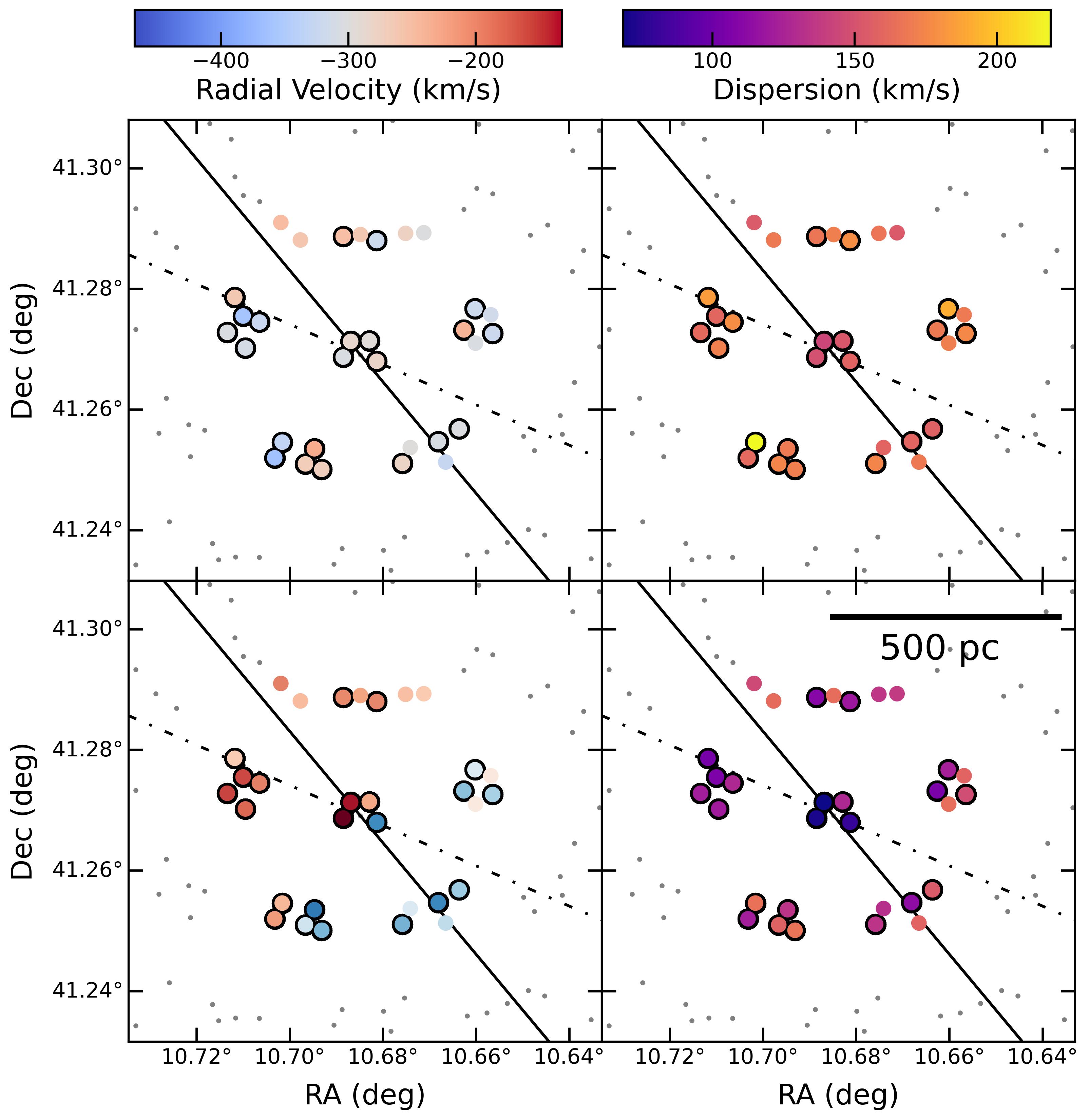}
\caption{Maps of the radial velocity (left column) and dispersion (right column) of Component 1 (top row) and 2 (bottom row) in the bulge region of M31. The 22 fiber positions that met the $\Delta$BIC cutoff have a thicker black outline. The disc and bar position angles are shown as full and dash-dotted lines, respectively. Other fiber positions in the region are shown as grey points. \label{fig:bulge_kins}}
\end{figure}

Figure~\ref{fig:bulge_kins} shows the results for each component's kinematics in the central bulge region of M31. Component 1 exhibits little, if any, overall rotation and has a velocity dispersion on average 38~\kmspace higher than Component 2. Component 2 shows highly structured rotation. Both components show an increase in velocity dispersion with distance from the centre, where distances have been deprojected into the plane of the disc of M31 as in \citetalias{Gibson+2023}. For Component 1 the dispersion gradient is $20.6^{+12.0}_{-11.7}$ \kmskpc, and for Component 2 it is $33.9^{+6.3}_{-6.5}$ \kmskpc.

\subsubsection{Abundance Gradients}

\begin{figure*}
\includegraphics[width=\textwidth]{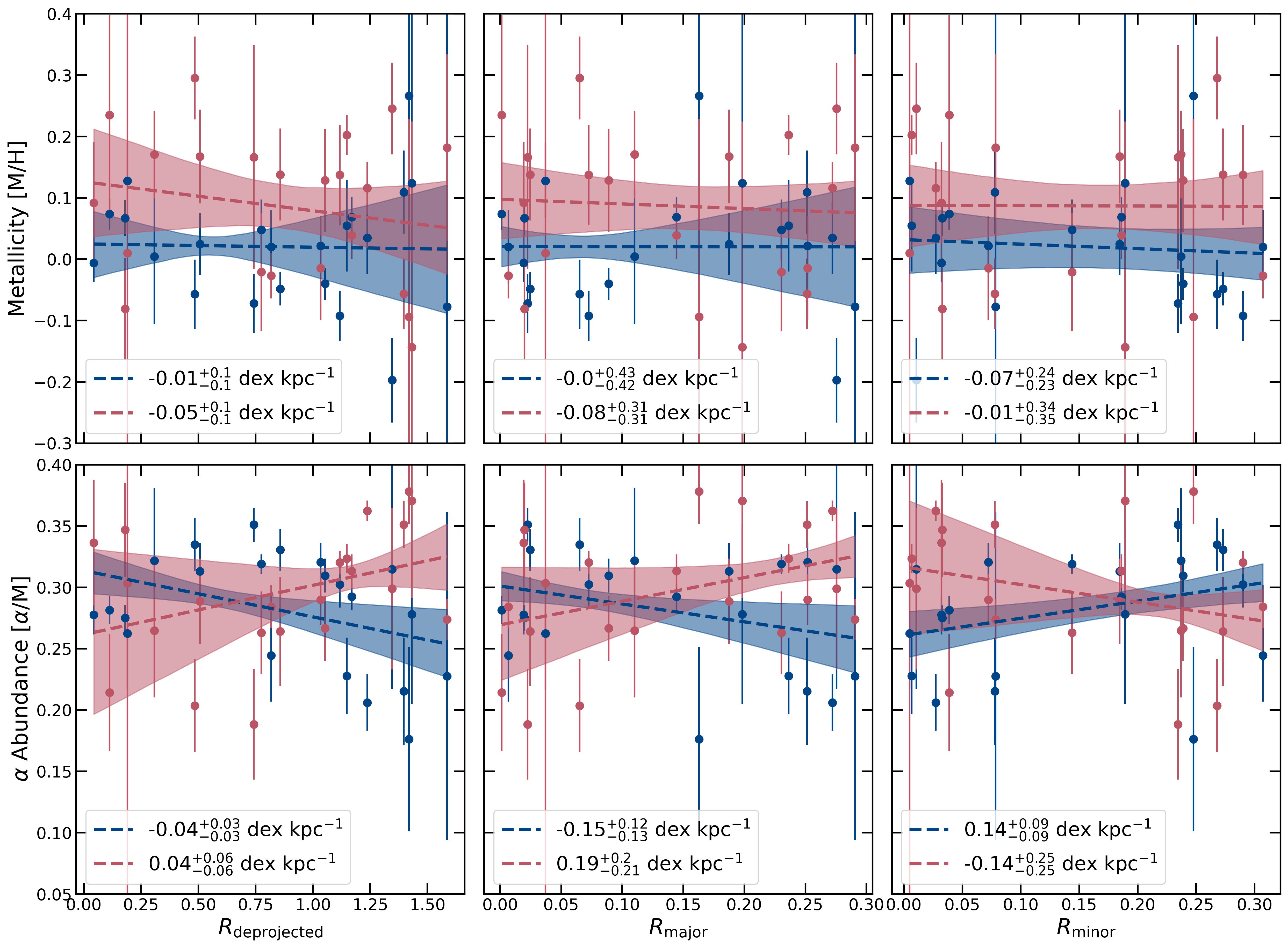}
\caption{Abundance gradients for Components 1 (blue) and 2 (red) in the bulge region of M31. The top row is for [M/H] and the bottom is for [$\alpha$/M], and the columns, from left to right, show gradients deprojected into the disc, projected perpendicular to the bar, and along the bar. Errors were measured using the jackknifing routine from Section~\ref{sec:bulgefits}. The shaded regions indicate the 16th and 84th percentiles of the slope measurements, and were calculated by bootstrapping the abundance measurements 5000 times. Gradients and their errors are written in each panel in units of dex kpc$^{-1}$. \label{fig:bulge_abunds}}
\end{figure*}

We examine the gradients in metallicity and $\alpha$ abundance for the two bulge components in Figure~\ref{fig:bulge_abunds}. Because it is unclear whether these gradients should follow the disc, bar, or just be radially distributed, we examine the gradients as a function of different quantities. The left column shows gradients with respect to the deprojected galactocentric radius ($R_{\rm deprojected}$) was calculating by deprojecting the on-sky distance into the plane of M31's disc, assuming an inclination angle of 77\degrees. We also present abundance gradients perpendicular to (middle column) and along the bar (right column). For this, we use the projected distance from the bar major and minor axes, $R_{\rm major}$ and $R_{\rm minor}$, respectively. With $R_{\rm major}$ we can determine how each component's chemistry changes from the on- and off-bar regions, and with $R_{\rm minor}$ we see the differences between the centre and the ends or ansae of the bar.

To calculate these gradients, we bootstrapped the measurements 5000 times using the jackknife-determined errors (Section~\ref{sec:bulgefits}). Quoted errors on the gradients are the 16th and 84th percentile slopes of the bootstrapped measurements, and are shown as the shaded regions in Figure~\ref{fig:bulge_abunds}.

As can be seen in Figure~\ref{fig:bulge_abunds}, there is little, if any, indication of a metallicity gradient in any direction for Component 1, while Component 2 has a slightly negative metallicity gradient when deprojected into the disc, though it exhibits little variation with respect to the bar. Both components exhibit steeper $\alpha$ gradients in all directions, and the signs of the [$\alpha$/M] gradients for the two populations are opposite. It is worth noting that none of our measurements are more than $\sim$1.5$\sigma$ from being flat, so we report no significant change in abundance in this small region of the galaxy.

Component 1 exhibits little overall rotation, high velocity dispersion, and effectively no metallicity gradient in any direction. It also has a slightly lower average metallicity than that of Component 2. Based on the following properties of component 1, we associate this with the classical bulge component of M31. Component 2 has very structured rotation and a lower velocity dispersion, as well as a slightly negative metallicity gradient in the disc and perpendicular to the bar, which is consistent with a bar. It is worth noting that Component 1 shows a strong negative $\alpha$ gradient in the disc and perpendicular to the bar, but a positive $\alpha$ gradient along the bar. The inverse of these trends are true, albeit less strongly, for Component 2. This indicates that there is a complex relationship between the bar position and the $\alpha$ abundance in this region of M31, and that the classical bulge is not uniformly enhanced, as one might expect.

\subsection{The disc}
\label{sec:discresults}

\begin{figure*}
\includegraphics[width=\textwidth]{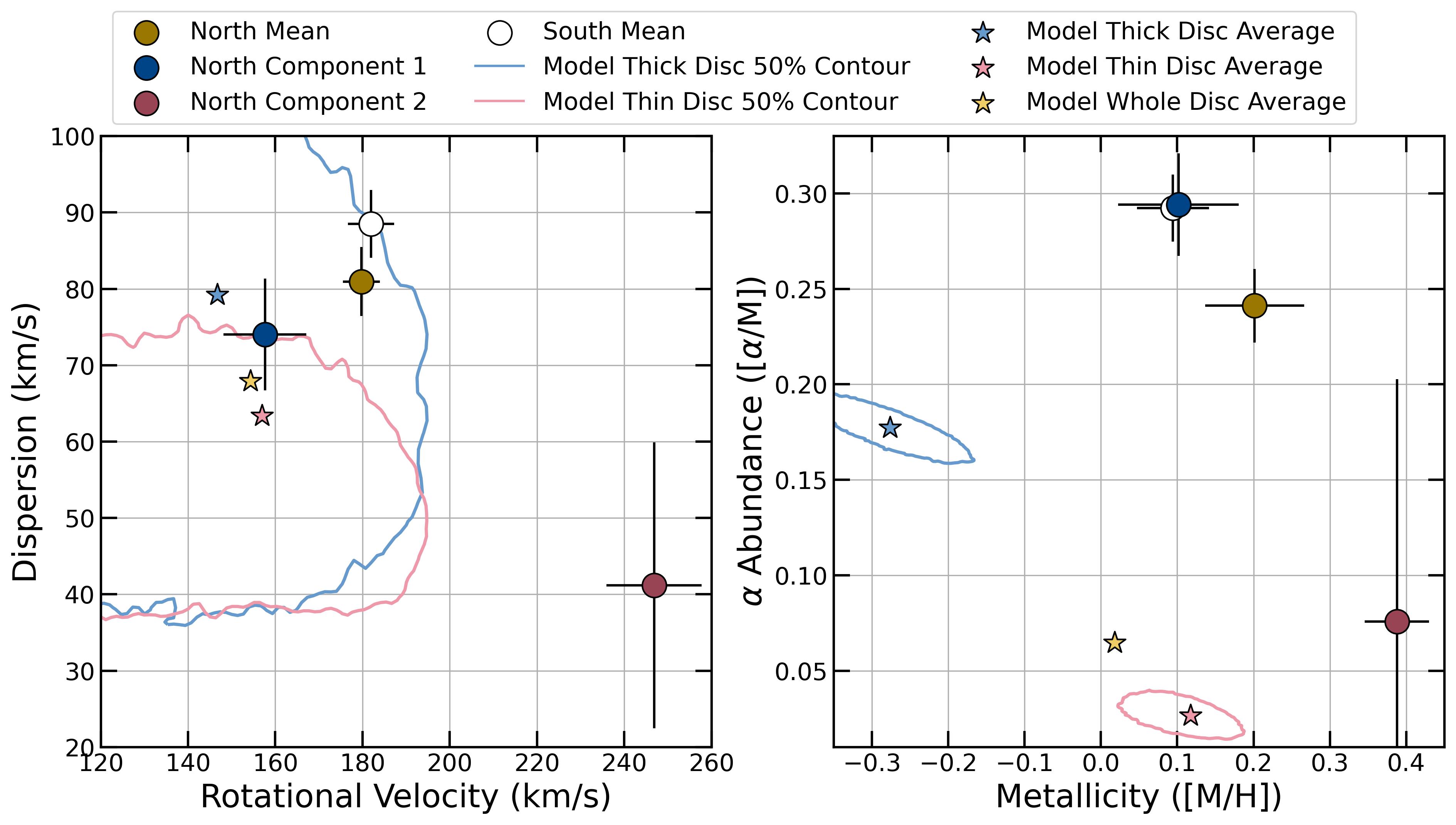}
\caption{Kinematical and chemical results from one-component fits in the northern (dark yellow) and southern (white) discs and from two-component (dark blue and dark red) fits in the northern disc. Errors on the measurements (black lines) come from jackknifing the observations, and individual jackknifed results are shown as coloured points. These results are plotted over chemodynamics from the E-Galaxia model MW disc from S. Sharma (in preparation), which has been projected at the location, distance, position angle, and inclination of M31 (see Section~\ref{sec:egalaxia}). Star points are the average values for the SSP particles within 1\arcsec of our north disc fiber positions; the contour indicates the 50\% enclosed level for all ssp particles. \label{fig:disc_results}}
\end{figure*}

In M31’s disc we focus on two regions in the north and south that span galactocentric radii of $\sim$4-7 kpc projected in the plane of the disc. We focus on these regions because the photometric decomposition from Section~\ref{sec:decomp} indicates $>80\%$ of the light here comes from the disc. We find clear evidence for two distinct chemodynamic components in the north, but only one in the south. The results from this analysis can be seen in Figure~\ref{fig:disc_results} and are listed in Table~\ref{tab:results}. The mean results in the northern and southern discs are shown as yellow and black points, respectively, in Figure~\ref{fig:disc_results}. Results for Components 1 and 2 in the north are shown as blue and red points, respectively; we do not present two-component results for the south, as the fits did not meet the $\Delta$BIC cutoff presented in Section~\ref{sec:deltabic}. In this figure we also compare our findings to a model MW disc (blue and red contours) that has been projected on-sky as M31 appears (see Section~\ref{sec:egalaxia} for details).

Component 1 in the north has a rotation velocity of 157.64 \kmspace and a velocity dispersion of 74.03 \kms, whereas Component 2 here rotates faster and more uniformly with a rotation velocity of 246.86 \kmspace and a velocity dispersion of 41.19 \kms. Component 1 is also more metal-poor and $\alpha$-rich ([M/H] = 0.10 and [$\alpha$/M] = 0.29) than Component 2 ([M/H] = 0.39 and [$\alpha$/M] = 0.08). We found the best fitting value for $f$ to be 0.8, which means Component 1 makes up 80\% of the total light emitted in the region, with Component 2 making up the other 20\%. The combination of kinematics and abundances derived here are consistent with Component 1 being an $\alpha$-rich thick disc and Component 2 being an $\alpha$-poor thin disc. We note that we have no information on the spatial structure of the two discs in this region, and therefore cannot constrain the relative thickness or thinness of the two components. Our use of ``thick" and ``thin" is simply to be analogous to the two kinematically- and chemically-defined disc components seen in the MW.

 The mean kinematics of the two sides of the disc are very similar, with a difference in rotational velocity of just 2.237~\kmspace and a difference in dispersion of 7.574~\kms. Their abundances are more distinct; however, the southern mean and northern Component 1 (blue points) abundances are very similar. One possible explanation is that the southern disc of M31 has been dynamically heated by interaction with M32 or the Giant Stellar Stream (GSS) and that this heating has not yet reached the northern disc. This would explain the slightly lower mean velocity dispersion in the north and the lack of an $\alpha$-poor thin disc component in the south. See Section~\ref{sec:litcomp_disc} for more discussion.

\subsection{Comparison to Literature and The Milky Way}
\label{sec:litcomp}

One of the high-level goals of this project is to assess not only the properties of M31 itself, but also its similarity and dissimilarity to the MW. Thus in this section, we not only compare our results to previous studies of M31 but also discuss them in the context of comparable properties of the MW.  

\subsubsection{M31's Bulge}

Our kinematics in the bulge are, on the whole, consistent with those measured by other studies. Our radial velocities in the bulge are centreed around -300 \kms, which is the systemic velocity of M31 \citep{M31velocity, M31Distance}. Averaging over all 22 bulge fiber positions, we find the one-component rotation velocity in the bulge of $\sim$28 \kms, which is broadly consistent with the trends found by \citet{Opitsch+2018}. Naturally, the rotation velocity we find for Component 2 in this region is larger, but given its smaller light-weighting fraction, $f$, this does not substantially affect the mean results to make them fall out of line with previous studies. The mean results for the velocity curve in the bulge well match our previ ous analysis in \citetalias{Gibson+2023}.

The one-component velocity dispersion in the bulge is $\sim$162~\kms. \citet{Opitsch+2018} finds the dispersion within the central 20" of M31 to be $\sim$160~\kms. They quote a mean dispersion for the whole bulge \citep[defined as where the bulge-to-total luminosity ratio is higher than 0.5 from the decomposition of][]{Kormendy+Bender1999} of $\sim$137 \kms, which is within the errors of our measurement for Component 1. We also measure a positive velocity dispersion gradient for both components in this region, which was seen by \citet{Opitsch+2018} and \citetalias{Gibson+2023}. 

\citet{Saglia+2018} analysed the same IFU data as \citet{Opitsch+2018} to characterize stellar abundances. They quantified the chemistry of the classical bulge, b/p bulge, and bar, and did so by isolating regions where the light is dominated by each component (see their Section~2 and Figure~3). They find that the inner regions of the classical bulge have an average metallicity of $\rm [Z/H] = 0.06$ and an $\alpha$ abundance of $\rm [\alpha/Fe] = 0.28$. This aligns best with our bulge Component 1, which we find to have $\rm [M/H] = 0.02$ and $\rm [\alpha/M] = 0.28$. 

They also define the region dominated by the bar to be about 1~kpc wide (deprojected into the disc) on either side of the inner classical bulge. They find this region to have [Z/H] = 0.02 and [$\alpha$/Fe] = 0.27. However they also note that this region has negative metallicity and $\alpha$ abundance gradients, so our finding of higher metallicity and $\alpha$ abundances in the more central regions of the bulge aligns reasonably well with their findings.

It is known that M31 has an old, massive classical bulge component \citep[e.g.][]{Athanassoula+Beaton2006, Saglia+2010}, however the MW does not. 50\% of the total stellar mass of the MW resides in its bulge \citep[][and references therein]{Licquia+Newman2015}, of which 60\% is associated with the bar and the other 40\% is associated with the inner disc \citep[e.g.][]{Portail+2017}. See also Section~4 of \citet{Bland-Hawthorn+Gerhard2016}. The bulge of the MW hosts a complex stellar population that spans $\sim$2.5 dex in metallicity \citep[e.g.][]{Zoccali+2008, Rojas-Arriagada+2019, Rojas-Arriagada+2020}. The kinematics of the stellar populations here are correlated with their chemistry; the stars on spheroidal orbits (more associated with the bulge) tend to be lower in metallicity, whereas the stars on ``bar-like" orbits tend to have higher metallicity \citep[e.g.,][but see also \citet{Bovy+2019}]{Hill+2011, Ness+2013, Ness+Lang2016, Zasowski+2016}. 

\subsubsection{M31's disc}
\label{sec:litcomp_disc}

In the northern part of the disc we identify and characterize two chemodynamically distinct stellar components, but but identify only a single one in the southern disc. We first note that in \citetalias{Gibson+2023} the stellar populations in the northern half exhibit a more varied mean chemistry than in the south (see their Figure~5).

\citet{Collins+2011} studied individual stellar spectra from Keck-DEIMOS in 21 fields in M31's southern disc. Their closest field to ours was 9.8 kpc from the centre along the disc major axis; our southern disc combined spectrum extends to roughly 5 kpc along the disc major axis. They split the 301 stars in this field into a thick and thin disc based on the stellar kinematics and studied the samples individually. Notably, their thin disc sample contained far more stars than their thick disc sample, and also had a higher velocity dispersion (55.2 \kmspace vs. 41.2 \kmspace for the thick disc). They found these stellar populations to have lower metallicities ([Fe/H] = -1.0 for the thick and -0.8 for the thin) than we see closer to the centre. It is known, however, that the disc of M31 has a negative dispersion and metallicity gradient beyond 4 kpc \citep[e.g.][]{PHAT_metal, Dorman+2015, Saglia+2018, Gibson+2023}, so it is possible that our metallicity results could align with those of \citet{Collins+2011}, but any further comparison is beyond the scope of this work. 

\citet{Nidever+2024} investigated JWST NIRSpec spectra of 103 RGB stars in the outer disc of M31. They found no strong evidence of a bimodal distribution of $\alpha$ abundance at constant metallicity. Instead, the chemistry of their stars lies along one sequence from $\rm [Fe/H] \simeq -0.4$ and $\rm [\alpha/Fe] \simeq +0.3$ to $\rm [Fe/H] \simeq +0.5$ and $\rm [\alpha/Fe] \simeq -0.2$, similar to the MW's thick disc.

\citet{Dorman+2015} combined Keck-DEIMOS spectra and PHAT photometry of 5800 stars in M31's northern disc to study the age-dispersion and age-metallicity relationships. They did so by splitting their sample into stars of different types (RGB, old/young AGB, and main sequence). They find a peak in the velocity dispersion distribution for RGB stars at $\sim$100 \kmspace; for older AGB stars this distribution peaks around 70 \kms. They investigate a region roughly 6 kpc from the M31's centre along the disc major axis that shows higher dispersion by a factor of 1.5-2. They attribute this feature to the superposition of two kinematic components: the disc and the end of the long bar. Additionally, they find the disc to be ``kinematically clumpy" due to a low velocity tail in the velocities of individual stars. That is, on scales smaller than about 0.75 kpc the kinematics are non-uniform, which indicates the presence of multiple overlapping kinematic components, similar to what we find in the same region.

In Section~\ref{sec:discresults} we posited that the non-identification of two chemodynamic components in the southern disc could be due to interaction with or contamination from M32 or the GSS. \citet{Dsouza+Bell2018} made a simulation of what an interaction between M31 and M32 may look like and determined that the M32 progenitor and M31 would have fully merged 2 Gyr ago, and that the remnant would be moving from the northeast to the northwest at present day. \citet{Hammer+2018} also found that the merger would have occured 2 Gyr ago. However \citet{Block+2006} and \citet{Gordon+2006} both find an interaction \lessthan200 Myr ago could create the ring/spiral arm nature of the disc. In between these two estimates, \citet{Dierickx+2014} modeled the same interaction and found that a collision $\sim$800~Myr ago and slightly off centre of M31 would produce the same structure. None of these studies go into great detail about potential differences between the two halves of the disc at present time. We hope that future analysis of the new Panchromatic Hubble Andromeda Southern Treasury (PHAST, Z. Chen, in preparation) will be able to further reconcile the differences between the two halves of Andromeda's disc.

All of these results combined paint a picture of M31's disc as a chemodynamically varied environment, likely with multiple overlapping kinematic components that have been mixed over time by mergers.

\subsubsection{The Milky Way's disc}
\label{sec:egalaxia}

The MW's disc is split into two  components that are kinematically and chemically different. One component is made up of stars that are generally older \citep[e.g.,][]{Haywood+2013, Martig+2016, Wu+2021}, more metal-poor, and more rich in $\alpha$ elements than those in the other component \citep[e.g.,][]{Fuhrmann1998, Bensby+2003, Bensby+2005, Reddy+2006, Haywood+2013, Anders+2014, Hayden+2015, Weinberg+2019, Imig+2023}. The former also has a larger scale height and shorter scale length than the latter \citep[e.g.][]{Yoshii1982, Gilmore+Reid1983}, and its stars are often on more eccentric orbits that take them farther beyond the midplane of the disc \citep[e.g.][]{Chiba+Beers2000, Bovy+2012c, Haywood+2013, Bovy+2016}. There is some debate as to how truly distinct these two components are \citep[e.g.,][]{Bovy+2012a}.

In order to obtain a more quantitative comparison of the two galaxies' disc we compare our results to a mock MW stellar catalog from \textsc{E-Galaxia} (S. Sharma, in preparation) in Figure~\ref{fig:disc_results}. \textsc{E-Galaxia} is a tool to generate SSP particles with 3-D position and velocity, age, [Fe/H], and [$\alpha$/Fe] based on the galactic chemical evolution (GCE) model of \citet{Sharma+2021b}. 
This GCE model includes a new prescription for the evolution of [$\alpha$/Fe] with age and [Fe/H], as well as a new set of stellar velocity dispersion relations from MW spectroscopic surveys \citep{Sharma+2021a}. Therefore, \textsc{E-Galaxia} is able to reproduce the [$\alpha$/Fe]-[Fe/H] distribution at different positions in the MW seen in APOGEE observations \citep{Hayden+2015}.

For this work we used the \textsc{E-Galaxia} mock catalog with $\sim10^8$ SSP particles and projected the model disc at M31's distance of 785~kpc, position angle of 39.1475\degrees, inclination of 77\degrees, and central RA and Dec of 10.6848\degreespace and 41.2691\degrees. We also scaled the size of the \textsc{E-Galaxia} disc by a factor of 1.8 to align with the scale radius of M31\footnote{We took the MW infrared scale radius of 2.6 kpc from \citet{Bland-Hawthorn+Gerhard2016} and the measured M31 scale radius from our photometric decomposition of 4.68 kpc.}. From this projection, we selected all SSP particles falling in the same region of the rescaled model galaxy as our northern disc sample of M31, using a cut in RA and Dec. The SSP particles were split into an $\alpha$-rich thick disc and an $\alpha$-poor thin disc sample using a cut on age at 10.5~Gyr, as this is defined to be the time of the sharp transition between the high- and low-$\alpha$ sequences \citep[see Section~5.1 of][]{Sharma+2021b}. We then isolated SSP particles within 1\arcsec (the radius of APOGEE fibers) of our fiber positions in the north disc and calculated the average rotation velocity, metallicity, and $\alpha$ abundance of the whole sample. The velocity dispersion was taken to be the standard deviation of the radial velocities of the sample. For the contours, we binned the SSP particles in each
sample into 2\arcsec square bins and calculated the mass-weighted average rotation velocity, metallicity, and $\alpha$ abundance for each bin. The velocity dispersion in was taken to be the standard deviation of the radial velocities in each bin, and we exclude bins with a dispersion of zero.

We note that the SSP particles of \textsc{E-Galaxia} each represent the same mass (1~M$_{\odot}$), so the values presented in Figure~\ref{fig:disc_results} are mass-weighted, as opposed to our M31 measurements, which are light-weighted. We are unable to straightforwardly convert between these two \citep[e.g.][]{Wang+2023}, but nevertheless this comparison is useful in a qualitative sense. The best-fitting light weighting fraction $f$ to our M31 northern disc spectrum was found to be 0.8, while the \textsc{E-Galaxia} thick-to-thin-disc mass weighting ratio for the whole region we selected is 0.335. This difference is large, but not unexpected, regardless of the weighting, since we know that the MW's thin disc dominates \citep[e.g.][]{Gilmore+Reid1983, Hayden+2015} and that M31's disc is thick and kinematically hot \citep[e.g.][]{Dorman+2015, PHAT_diskthick}.

We compare our M31 results to the \textsc{E-Galaxia} model disc in Figure~\ref{fig:disc_results}. Our M31 results are shown as coloured circles, and the model disc is represented with contours enclosing 50\% of the SSP particles for each component, as well as stars showing the mass-weighted average value for each component and the disc as a whole.
We emphasize that this comparison is to highlight relative trends between different galactic components, since \textsc{E-Galaxia} is tuned specifically to the MW and not M31.

In the left panel we see that the kinematics of the M31 and model $\alpha$-rich thick discs (blue) are reasonably well aligned --- the average rotational velocity and dispersion of SSP particles in our fiber positions is just 10.95 \kmspace lower and 5.21 \kmspace higher, respectively, than our north Component 1 measurements. However, the $\alpha$-poor thin disc kinematics differ substantially, with the \textsc{E-Galaxia} model thin disc having a much slower rotation velocity and higher dispersion. Light-weighted velocity dispersions should be lower than mass-weighted ones since young populations have a lower dispersion but contribute more to the luminosity than they do the mass, though the amount can vary based on projection and other effects. Also, higher-dispersion components will naturally exhibit a slower average rotation velocity (as discussed in Section~\ref{sec:twocomp}), but the existence of such a rapidly-rotating component in M31's disc is at the very least intriguing.

The chemistry of M31 and the \textsc{E-Galaxia} disc are compared in the right panel of Figure~\ref{fig:disc_results}. The \textsc{E-Galaxia} metallicities and $\alpha$ abundances are both lower than what we find for M31, but the {\it difference} in abundance between the two components in each galaxy is similar. For \textsc{E-Galaxia} the two component's average metallicities and $\alpha$ abundances differ by 0.39 and 0.15~dex, respectively. For M31 the differences are 0.29 and 0.22~dex. Additionally, we see that the mean (one-component) abundances for M31 and the whole disc average abundances for \textsc{E-Galaxia} fall directly between the values for each component. In fact, if we calculate the mass- or light-weighted average of the two component's abundances, they align very accurately with the mean and whole disc result\footnote{So for the M31 $\alpha$ abundances, we can calculate $f{\rm [\alpha/M]}_1 + (1-f) {\rm [\alpha/M]}_2 = 0.2505$, which differs by less than 0.01 dex from $\rm [\alpha/M]_{mean}$.}. 

Thus, we identify two different chemodynamic components in the disc of M31 which have a comparable spread, though offset, in average abundances to those seen in the MW. It is also evident that M31's disc is much is more dominated by it's $\alpha$-rich thick component than the MW's is. In the MW the majority of its stars in a comparable region are associated with the $\alpha$-poor thin disc. This potentially points to different evolutionary paths for the two galaxies, which is unsurprising given their rather different merger histories.

\section{Summary and Conclusions}
\label{sec:summary}
We present an analysis of high-resolution, integrated light, NIR APOGEE spectra in the bulge and inner disc of M31. We studied 31 spectra in the inner 91.5\arcsec\space ($\sim$0.35~kpc) of the bulge as well as two co-added spectra in the north and south of the disc ($R \sim 4-7$~kpc). We fit all 33 spectra with both one and two SSP spectral templates, using physically motivated kinematics priors on each component, as well as a photometric decomposition for the relative light-weighting of each component in the two-template fit. We rigorously tested our methodology using realistic mock observations to ensure we reliably recovered input parameters. We used the Bayesian Information Criterion to assess whether or not using two components produced better fits than one component. Our final results are presented in Table~\ref{tab:results} and Figures \ref{fig:bulge_kins}, \ref{fig:bulge_abunds}, and \ref{fig:disc_results}. Our takeaways are summarised here.
\begin{itemize}
    \item In 22 of our 31 bulge spectra, we identified two distinct chemodynamic components. Component 1 exhibits little overall rotation and a high velocity dispersion. Component 2 exhibits highly structured rotation and a lower average velocity dispersion. Neither component shows much indication of a metallicity gradient deprojected into the disc or along or perpendicular to the bar; however, both components do seem to have $\alpha$ abundances that are anti-correlated in all three directions, indicating a complex abundance relationship with the bar position.
    \item In the northern disc spectrum, we identified two distinct chemodynamic components. Component 1 rotates slower and has a higher velocity dispersion, lower metallicity, and higher $\alpha$ abundance than Component 2. We found that Component 1 makes up 80\% of the light coming from this region.
    \item We did not identify two distinct chemodynamic components in the southern disc. Instead, a better fit was provided by the one-component result, which is kinematically very similar to the mean (one-component) result in the north, and chemically similar to Component 1 in the north. 
\end{itemize}

We have identified and characterized two overlapping stellar populations in the bulge and northern disc of M31. In the bulge, Component 1 is chemodynamically consistent with what we expect from a classical bulge, while Component 2 is consistent with a bar. In the northern disc, we identified and characterized patterns consistent with a dominant $\alpha$-poor thick disc and a $\alpha$-rich thin disc, {\it for the first time associating the spread in M31's $\alpha$ abundance with kinematically distinct stellar populations}. We find the southern disc to be chemodynamically similar to the northern $\alpha$-rich thick disc, but lacking a distinct $\alpha$-poor thin disc component. 

If this narrative interpretation is correct, it suggests that the complex patterns seen in the MW's stars are not unique in the universe, but also that galactic asymmetries can further complicated the picture. Such analysis on more galaxies in the future will continue to constrain the processes that govern the structural formation and evolution of galaxies.

\section*{Acknowledgements}

BJG, GZ, and AS acknowledge support for this project provided by NSF grant AST-1911129. DAG is supported by STFC grants ST/T000244/1 and ST/X001075/1. 

Funding for the Sloan Digital Sky Survey IV has been provided by the Alfred P. Sloan Foundation, the U.S. Department of Energy Office of Science, and the Participating Institutions. SDSS-IV acknowledges
support and resources from the Center for High-Performance Computing at
the University of Utah. The SDSS web site is www.sdss.org.

SDSS-IV is managed by the Astrophysical Research Consortium for the 
Participating Institutions of the SDSS Collaboration including the 
Brazilian Participation Group, the Carnegie Institution for Science, 
Carnegie Mellon University, the Chilean Participation Group, the French Participation Group, Harvard-Smithsonian Center for Astrophysics, 
Instituto de Astrof\'isica de Canarias, The Johns Hopkins University, Kavli Institute for the Physics and Mathematics of the Universe (IPMU) / 
University of Tokyo, the Korean Participation Group, Lawrence Berkeley National Laboratory, 
Leibniz Institut f\"ur Astrophysik Potsdam (AIP), 
Max-Planck-Institut f\"ur Astronomie (MPIA Heidelberg), 
Max-Planck-Institut f\"ur Astrophysik (MPA Garching), 
Max-Planck-Institut f\"ur Extraterrestrische Physik (MPE), 
National Astronomical Observatories of China, New Mexico State University, 
New York University, University of Notre Dame, 
Observat\'ario Nacional / MCTI, The Ohio State University, 
Pennsylvania State University, Shanghai Astronomical Observatory, 
United Kingdom Participation Group,
Universidad Nacional Aut\'onoma de M\'exico, University of Arizona, 
University of Colorado Boulder, University of Oxford, University of Portsmouth, 
University of Utah, University of Virginia, University of Washington, University of Wisconsin, 
Vanderbilt University, and Yale University.

The Digitized Sky Survey was produced at the Space Telescope Science Institute under U.S. Government grant NAG W-2166. The images of these surveys are based on photographic data obtained using the Oschin Schmidt Telescope on Palomar Mountain and the UK Schmidt Telescope. The plates were processed into the present compressed digital form with the permission of these institutions.

This work has made use of unWISE images (DOI: 10.26131/IRSA524). unWISE makes use of data products from the Wide-field Infrared Survey Explorer, which is a joint project of the University of California, Los Angeles, and the Jet Propulsion Laboratory/California Institute of Technology, funded by the National Aeronautics and Space Administration.

\section*{Data Availability}

The APOGEE DR17 data underlying this article are described (including access details) here: \url{https://www.sdss4.org/dr17/irspec/}. 
The custom stacked spectra generated from these data may be shared on reasonable request to the corresponding author.



\bibliographystyle{mnras}
\bibliography{main} 

\begin{thebibliography}{}
\makeatletter
\relax
\def\mn@urlcharsother{\let\do\@makeother \do\$\do\&\do\#\do\^\do\_\do\%\do\~}
\def\mn@doi{\begingroup\mn@urlcharsother \@ifnextchar [ {\mn@doi@} {\mn@doi@[]}}
\def\mn@doi@[#1]#2{\def\@tempa{#1}\ifx\@tempa\@empty \href {http://dx.doi.org/#2} {doi:#2}\else \href {http://dx.doi.org/#2} {#1}\fi \endgroup}
\def\mn@eprint#1#2{\mn@eprint@#1:#2::\@nil}
\def\mn@eprint@arXiv#1{\href {http://arxiv.org/abs/#1} {{\tt arXiv:#1}}}
\def\mn@eprint@dblp#1{\href {http://dblp.uni-trier.de/rec/bibtex/#1.xml} {dblp:#1}}
\def\mn@eprint@#1:#2:#3:#4\@nil{\def\@tempa {#1}\def\@tempb {#2}\def\@tempc {#3}\ifx \@tempc \@empty \let \@tempc \@tempb \let \@tempb \@tempa \fi \ifx \@tempb \@empty \def\@tempb {arXiv}\fi \@ifundefined {mn@eprint@\@tempb}{\@tempb:\@tempc}{\expandafter \expandafter \csname mn@eprint@\@tempb\endcsname \expandafter{\@tempc}}}

\bibitem[\protect\citeauthoryear{{Abdurro'uf} et~al.,}{{Abdurro'uf} et~al.}{2022}]{Abdurro'uf+2022}
{Abdurro'uf} et~al., 2022, \mn@doi [\apjs] {10.3847/1538-4365/ac4414}, \href {https://ui.adsabs.harvard.edu/abs/2022ApJS..259...35A} {259, 35}

\bibitem[\protect\citeauthoryear{{Almeida} et~al.,}{{Almeida} et~al.}{2023}]{Almeida_2023_dr18}
{Almeida} A.,  et~al., 2023, \apjs, 267, 44

\bibitem[\protect\citeauthoryear{{Anders} et~al.,}{{Anders} et~al.}{2014}]{Anders+2014}
{Anders} F.,  et~al., 2014, \mn@doi [\aap] {10.1051/0004-6361/201323038}, \href {https://ui.adsabs.harvard.edu/abs/2014A&A...564A.115A} {564, A115}

\bibitem[\protect\citeauthoryear{{Ashok} et~al.,}{{Ashok} et~al.}{2021}]{Ashok+2021}
{Ashok} A.,  et~al., 2021, \mn@doi [\aj] {10.3847/1538-3881/abd7f1}, \href {https://ui.adsabs.harvard.edu/abs/2021AJ....161..167A} {161, 167}

\bibitem[\protect\citeauthoryear{{Athanassoula} \& {Beaton}}{{Athanassoula} \& {Beaton}}{2006}]{Athanassoula+Beaton2006}
{Athanassoula} E.,  {Beaton} R.~L.,  2006, \mn@doi [\mnras] {10.1111/j.1365-2966.2006.10567.x}, \href {https://ui.adsabs.harvard.edu/abs/2006MNRAS.370.1499A} {370, 1499}

\bibitem[\protect\citeauthoryear{{Barmby} et~al.,}{{Barmby} et~al.}{2006}]{Barmby+2006}
{Barmby} P.,  et~al., 2006, \mn@doi [\apjl] {10.1086/508626}, \href {https://ui.adsabs.harvard.edu/abs/2006ApJ...650L..45B} {650, L45}

\bibitem[\protect\citeauthoryear{{Beaton} et~al.,}{{Beaton} et~al.}{2007}]{Beaton+2007}
{Beaton} R.~L.,  et~al., 2007, \mn@doi [\apjl] {10.1086/514333}, \href {https://ui.adsabs.harvard.edu/abs/2007ApJ...658L..91B} {658, L91}

\bibitem[\protect\citeauthoryear{{Beaton} et~al.,}{{Beaton} et~al.}{2021}]{Beaton+2021}
{Beaton} R.~L.,  et~al., 2021, \mn@doi [\aj] {10.3847/1538-3881/ac260c}, \href {https://ui.adsabs.harvard.edu/abs/2021AJ....162..302B} {162, 302}

\bibitem[\protect\citeauthoryear{{Bensby}, {Feltzing}  \& {Lundstr{\"o}m}}{{Bensby} et~al.}{2003}]{Bensby+2003}
{Bensby} T.,  {Feltzing} S.,   {Lundstr{\"o}m} I.,  2003, \mn@doi [\aap] {10.1051/0004-6361:20031213}, \href {https://ui.adsabs.harvard.edu/abs/2003A&A...410..527B} {410, 527}

\bibitem[\protect\citeauthoryear{{Bensby}, {Feltzing}, {Lundstr{\"o}m}  \& {Ilyin}}{{Bensby} et~al.}{2005}]{Bensby+2005}
{Bensby} T.,  {Feltzing} S.,  {Lundstr{\"o}m} I.,   {Ilyin} I.,  2005, \mn@doi [\aap] {10.1051/0004-6361:20040332}, \href {https://ui.adsabs.harvard.edu/abs/2005A&A...433..185B} {433, 185}

\bibitem[\protect\citeauthoryear{{Bland-Hawthorn} \& {Gerhard}}{{Bland-Hawthorn} \& {Gerhard}}{2016}]{Bland-Hawthorn+Gerhard2016}
{Bland-Hawthorn} J.,  {Gerhard} O.,  2016, \mn@doi [\araa] {10.1146/annurev-astro-081915-023441}, \href {https://ui.adsabs.harvard.edu/abs/2016ARA&A..54..529B} {54, 529}

\bibitem[\protect\citeauthoryear{{Blanton} et~al.,}{{Blanton} et~al.}{2017}]{Blanton+2017}
{Blanton} M.~R.,  et~al., 2017, \mn@doi [\aj] {10.3847/1538-3881/aa7567}, \href {https://ui.adsabs.harvard.edu/abs/2017AJ....154...28B} {154, 28}

\bibitem[\protect\citeauthoryear{{Block} et~al.,}{{Block} et~al.}{2006}]{Block+2006}
{Block} D.~L.,  et~al., 2006, \mn@doi [\nat] {10.1038/nature05184}, \href {https://ui.adsabs.harvard.edu/abs/2006Natur.443..832B} {443, 832}

\bibitem[\protect\citeauthoryear{{Bournaud}, {Elmegreen}  \& {Martig}}{{Bournaud} et~al.}{2009}]{Bournaud+2009}
{Bournaud} F.,  {Elmegreen} B.~G.,   {Martig} M.,  2009, \mn@doi [\apjl] {10.1088/0004-637X/707/1/L1}, \href {https://ui.adsabs.harvard.edu/abs/2009ApJ...707L...1B} {707, L1}

\bibitem[\protect\citeauthoryear{{Bovy}, {Rix}  \& {Hogg}}{{Bovy} et~al.}{2012a}]{Bovy+2012a}
{Bovy} J.,  {Rix} H.-W.,   {Hogg} D.~W.,  2012a, \mn@doi [\apj] {10.1088/0004-637X/751/2/131}, \href {https://ui.adsabs.harvard.edu/abs/2012ApJ...751..131B} {751, 131}

\bibitem[\protect\citeauthoryear{{Bovy}, {Rix}, {Hogg}, {Beers}, {Lee}  \& {Zhang}}{{Bovy} et~al.}{2012b}]{Bovy+2012c}
{Bovy} J.,  {Rix} H.-W.,  {Hogg} D.~W.,  {Beers} T.~C.,  {Lee} Y.~S.,   {Zhang} L.,  2012b, \mn@doi [\apj] {10.1088/0004-637X/755/2/115}, \href {https://ui.adsabs.harvard.edu/abs/2012ApJ...755..115B} {755, 115}

\bibitem[\protect\citeauthoryear{{Bovy}, {Rix}, {Schlafly}, {Nidever}, {Holtzman}, {Shetrone}  \& {Beers}}{{Bovy} et~al.}{2016}]{Bovy+2016}
{Bovy} J.,  {Rix} H.-W.,  {Schlafly} E.~F.,  {Nidever} D.~L.,  {Holtzman} J.~A.,  {Shetrone} M.,   {Beers} T.~C.,  2016, \mn@doi [\apj] {10.3847/0004-637X/823/1/30}, \href {https://ui.adsabs.harvard.edu/abs/2016ApJ...823...30B} {823, 30}

\bibitem[\protect\citeauthoryear{{Bovy}, {Leung}, {Hunt}, {Mackereth}, {Garc{\'\i}a-Hern{\'a}ndez}  \& {Roman-Lopes}}{{Bovy} et~al.}{2019}]{Bovy+2019}
{Bovy} J.,  {Leung} H.~W.,  {Hunt} J. A.~S.,  {Mackereth} J.~T.,  {Garc{\'\i}a-Hern{\'a}ndez} D.~A.,   {Roman-Lopes} A.,  2019, \mn@doi [\mnras] {10.1093/mnras/stz2891}, \href {https://ui.adsabs.harvard.edu/abs/2019MNRAS.490.4740B} {490, 4740}

\bibitem[\protect\citeauthoryear{{Bowen} \& {Vaughan}}{{Bowen} \& {Vaughan}}{1973}]{Bowen+Vaughan1973}
{Bowen} I.~S.,  {Vaughan} A.~H. J.,  1973, \mn@doi [\ao] {10.1364/AO.12.001430}, \href {https://ui.adsabs.harvard.edu/abs/1973ApOpt..12.1430B} {12, 1430}

\bibitem[\protect\citeauthoryear{{Cappellari}}{{Cappellari}}{2023}]{pPXF}
{Cappellari} M.,  2023, \mn@doi [MNRAS] {10.1093/mnras/stad2597}, 526, 3273

\bibitem[\protect\citeauthoryear{{Casey}, {Hogg}, {Ness}, {Rix}, {Ho}  \& {Gilmore}}{{Casey} et~al.}{2016}]{TheCannon2}
{Casey} A.~R.,  {Hogg} D.~W.,  {Ness} M.,  {Rix} H.-W.,  {Ho} A. Q.~Y.,   {Gilmore} G.,  2016, arXiv e-prints, \href {https://ui.adsabs.harvard.edu/abs/2016arXiv160303040C} {p. arXiv:1603.03040}

\bibitem[\protect\citeauthoryear{{Chiappini}, {Matteucci}  \& {Gratton}}{{Chiappini} et~al.}{1997}]{Chiappini+1997}
{Chiappini} C.,  {Matteucci} F.,   {Gratton} R.,  1997, \mn@doi [\apj] {10.1086/303726}, \href {https://ui.adsabs.harvard.edu/abs/1997ApJ...477..765C} {477, 765}

\bibitem[\protect\citeauthoryear{{Chiappini}, {Matteucci}  \& {Romano}}{{Chiappini} et~al.}{2001}]{Chiappini+2001}
{Chiappini} C.,  {Matteucci} F.,   {Romano} D.,  2001, \mn@doi [\apj] {10.1086/321427}, \href {https://ui.adsabs.harvard.edu/abs/2001ApJ...554.1044C} {554, 1044}

\bibitem[\protect\citeauthoryear{{Chiba} \& {Beers}}{{Chiba} \& {Beers}}{2000}]{Chiba+Beers2000}
{Chiba} M.,  {Beers} T.~C.,  2000, \mn@doi [\aj] {10.1086/301409}, \href {https://ui.adsabs.harvard.edu/abs/2000AJ....119.2843C} {119, 2843}

\bibitem[\protect\citeauthoryear{{Clarke} et~al.,}{{Clarke} et~al.}{2019}]{Clarke+2019}
{Clarke} A.~J.,  et~al., 2019, \mn@doi [\mnras] {10.1093/mnras/stz104}, \href {https://ui.adsabs.harvard.edu/abs/2019MNRAS.484.3476C} {484, 3476}

\bibitem[\protect\citeauthoryear{{Collins} et~al.,}{{Collins} et~al.}{2011}]{Collins+2011}
{Collins} M.~L.~M.,  et~al., 2011, \mn@doi [\mnras] {10.1111/j.1365-2966.2011.18238.x}, \href {https://ui.adsabs.harvard.edu/abs/2011MNRAS.413.1548C} {413, 1548}

\bibitem[\protect\citeauthoryear{{Cui} et~al.,}{{Cui} et~al.}{2012}]{LAMOST}
{Cui} X.-Q.,  et~al., 2012, \mn@doi [Research in Astronomy and Astrophysics] {10.1088/1674-4527/12/9/003}, \href {https://ui.adsabs.harvard.edu/abs/2012RAA....12.1197C} {12, 1197}

\bibitem[\protect\citeauthoryear{{D'Souza} \& {Bell}}{{D'Souza} \& {Bell}}{2018}]{Dsouza+Bell2018}
{D'Souza} R.,  {Bell} E.~F.,  2018, \mn@doi [Nature Astronomy] {10.1038/s41550-018-0533-x}, \href {https://ui.adsabs.harvard.edu/abs/2018NatAs...2..737D} {2, 737}

\bibitem[\protect\citeauthoryear{{Dalcanton} et~al.,}{{Dalcanton} et~al.}{2012}]{PHAT}
{Dalcanton} J.~J.,  et~al., 2012, \mn@doi [\apjs] {10.1088/0067-0049/200/2/18}, \href {https://ui.adsabs.harvard.edu/abs/2012ApJS..200...18D} {200, 18}

\bibitem[\protect\citeauthoryear{{Dalcanton} et~al.,}{{Dalcanton} et~al.}{2023}]{PHAT_diskthick}
{Dalcanton} J.~J.,  et~al., 2023, \mn@doi [\aj] {10.3847/1538-3881/accc83}, \href {https://ui.adsabs.harvard.edu/abs/2023AJ....166...80D} {166, 80}

\bibitem[\protect\citeauthoryear{{De Silva} et~al.,}{{De Silva} et~al.}{2015}]{GALAH}
{De Silva} G.~M.,  et~al., 2015, \mn@doi [\mnras] {10.1093/mnras/stv327}, \href {https://ui.adsabs.harvard.edu/abs/2015MNRAS.449.2604D} {449, 2604}

\bibitem[\protect\citeauthoryear{{Dierickx}, {Blecha}  \& {Loeb}}{{Dierickx} et~al.}{2014}]{Dierickx+2014}
{Dierickx} M.,  {Blecha} L.,   {Loeb} A.,  2014, \mn@doi [\apjl] {10.1088/2041-8205/788/2/L38}, \href {https://ui.adsabs.harvard.edu/abs/2014ApJ...788L..38D} {788, L38}

\bibitem[\protect\citeauthoryear{{Dorman} et~al.,}{{Dorman} et~al.}{2015}]{Dorman+2015}
{Dorman} C.~E.,  et~al., 2015, \mn@doi [\apj] {10.1088/0004-637X/803/1/24}, \href {https://ui.adsabs.harvard.edu/abs/2015ApJ...803...24D} {803, 24}

\bibitem[\protect\citeauthoryear{{Eisenstein} et~al.,}{{Eisenstein} et~al.}{2011}]{Eisenstein+2011}
{Eisenstein} D.~J.,  et~al., 2011, \mn@doi [\aj] {10.1088/0004-6256/142/3/72}, \href {https://ui.adsabs.harvard.edu/abs/2011AJ....142...72E} {142, 72}

\bibitem[\protect\citeauthoryear{{Erwin}}{{Erwin}}{2015}]{IMFIT}
{Erwin} P.,  2015, \mn@doi [\apj] {10.1088/0004-637X/799/2/226}, \href {https://ui.adsabs.harvard.edu/abs/2015ApJ...799..226E} {799, 226}

\bibitem[\protect\citeauthoryear{{Fardal}, {Babul}, {Geehan}  \& {Guhathakurta}}{{Fardal} et~al.}{2006}]{Fardal+2006}
{Fardal} M.~A.,  {Babul} A.,  {Geehan} J.~J.,   {Guhathakurta} P.,  2006, \mn@doi [\mnras] {10.1111/j.1365-2966.2005.09864.x}, \href {https://ui.adsabs.harvard.edu/abs/2006MNRAS.366.1012F} {366, 1012}

\bibitem[\protect\citeauthoryear{{Fardal}, {Babul}, {Guhathakurta}, {Gilbert}  \& {Dodge}}{{Fardal} et~al.}{2008}]{Fardal+2008}
{Fardal} M.~A.,  {Babul} A.,  {Guhathakurta} P.,  {Gilbert} K.~M.,   {Dodge} C.,  2008, \mn@doi [\apjl] {10.1086/590386}, \href {https://ui.adsabs.harvard.edu/abs/2008ApJ...682L..33F} {682, L33}

\bibitem[\protect\citeauthoryear{{Ferguson} \& {Mackey}}{{Ferguson} \& {Mackey}}{2016}]{Ferguson+Mackey2016}
{Ferguson} A. M.~N.,  {Mackey} A.~D.,  2016, in {Newberg} H.~J.,  {Carlin} J.~L.,  eds,  Astrophysics and Space Science Library Vol. 420, Tidal Streams in the Local Group and Beyond. p.~191 (\mn@eprint {arXiv} {1603.01993}), \mn@doi{10.1007/978-3-319-19336-6_8}

\bibitem[\protect\citeauthoryear{{Font}, {Johnston}, {Guhathakurta}, {Majewski}  \& {Rich}}{{Font} et~al.}{2006}]{Font+2006}
{Font} A.~S.,  {Johnston} K.~V.,  {Guhathakurta} P.,  {Majewski} S.~R.,   {Rich} R.~M.,  2006, \mn@doi [\aj] {10.1086/499564}, \href {https://ui.adsabs.harvard.edu/abs/2006AJ....131.1436F} {131, 1436}

\bibitem[\protect\citeauthoryear{{Foreman-Mackey}, {Hogg}, {Lang}  \& {Goodman}}{{Foreman-Mackey} et~al.}{2013}]{emcee}
{Foreman-Mackey} D.,  {Hogg} D.~W.,  {Lang} D.,   {Goodman} J.,  2013, \mn@doi [\pasp] {10.1086/670067}, \href {https://ui.adsabs.harvard.edu/abs/2013PASP..125..306F} {125, 306}

\bibitem[\protect\citeauthoryear{{Fuhrmann}}{{Fuhrmann}}{1998}]{Fuhrmann1998}
{Fuhrmann} K.,  1998, \aap, \href {https://ui.adsabs.harvard.edu/abs/1998A&A...338..161F} {338, 161}

\bibitem[\protect\citeauthoryear{{Gaia Collaboration} et~al.,}{{Gaia Collaboration} et~al.}{2016}]{Gaia_mission}
{Gaia Collaboration} et~al., 2016, \mn@doi [\aap] {10.1051/0004-6361/201629272}, \href {https://ui.adsabs.harvard.edu/abs/2016A&A...595A...1G} {595, A1}

\bibitem[\protect\citeauthoryear{{Garc{\'\i}a P{\'e}rez} et~al.,}{{Garc{\'\i}a P{\'e}rez} et~al.}{2016}]{ASPCAP}
{Garc{\'\i}a P{\'e}rez} A.~E.,  et~al., 2016, \mn@doi [\aj] {10.3847/0004-6256/151/6/144}, \href {https://ui.adsabs.harvard.edu/abs/2016AJ....151..144G} {151, 144}

\bibitem[\protect\citeauthoryear{{Gibson} et~al.,}{{Gibson} et~al.}{2023}]{Gibson+2023}
{Gibson} B.~J.,  et~al., 2023, \apj, 952

\bibitem[\protect\citeauthoryear{{Gilmore} \& {Reid}}{{Gilmore} \& {Reid}}{1983}]{Gilmore+Reid1983}
{Gilmore} G.,  {Reid} N.,  1983, \mn@doi [\mnras] {10.1093/mnras/202.4.1025}, \href {https://ui.adsabs.harvard.edu/abs/1983MNRAS.202.1025G} {202, 1025}

\bibitem[\protect\citeauthoryear{{Gordon} et~al.,}{{Gordon} et~al.}{2006}]{Gordon+2006}
{Gordon} K.~D.,  et~al., 2006, \mn@doi [\apjl] {10.1086/501046}, \href {https://ui.adsabs.harvard.edu/abs/2006ApJ...638L..87G} {638, L87}

\bibitem[\protect\citeauthoryear{{Gregersen} et~al.,}{{Gregersen} et~al.}{2015}]{PHAT_metal}
{Gregersen} D.,  et~al., 2015, \mn@doi [\aj] {10.1088/0004-6256/150/6/189}, \href {https://ui.adsabs.harvard.edu/abs/2015AJ....150..189G} {150, 189}

\bibitem[\protect\citeauthoryear{{Gunn} et~al.,}{{Gunn} et~al.}{2006}]{Gunn+2006}
{Gunn} J.~E.,  et~al., 2006, \mn@doi [\aj] {10.1086/500975}, \href {https://ui.adsabs.harvard.edu/abs/2006AJ....131.2332G} {131, 2332}

\bibitem[\protect\citeauthoryear{{Hammer}, {Puech}, {Chemin}, {Flores}  \& {Lehnert}}{{Hammer} et~al.}{2007}]{Hammer+2007}
{Hammer} F.,  {Puech} M.,  {Chemin} L.,  {Flores} H.,   {Lehnert} M.~D.,  2007, \mn@doi [\apj] {10.1086/516727}, \href {https://ui.adsabs.harvard.edu/abs/2007ApJ...662..322H} {662, 322}

\bibitem[\protect\citeauthoryear{{Hammer}, {Yang}, {Wang}, {Ibata}, {Flores}  \& {Puech}}{{Hammer} et~al.}{2018}]{Hammer+2018}
{Hammer} F.,  {Yang} Y.~B.,  {Wang} J.~L.,  {Ibata} R.,  {Flores} H.,   {Puech} M.,  2018, \mn@doi [\mnras] {10.1093/mnras/stx3343}, \href {https://ui.adsabs.harvard.edu/abs/2018MNRAS.475.2754H} {475, 2754}

\bibitem[\protect\citeauthoryear{{Hayden} et~al.,}{{Hayden} et~al.}{2015}]{Hayden+2015}
{Hayden} M.~R.,  et~al., 2015, \mn@doi [\apj] {10.1088/0004-637X/808/2/132}, \href {https://ui.adsabs.harvard.edu/abs/2015ApJ...808..132H} {808, 132}

\bibitem[\protect\citeauthoryear{{Haywood}, {Di Matteo}, {Lehnert}, {Katz}  \& {G{\'o}mez}}{{Haywood} et~al.}{2013}]{Haywood+2013}
{Haywood} M.,  {Di Matteo} P.,  {Lehnert} M.~D.,  {Katz} D.,   {G{\'o}mez} A.,  2013, \mn@doi [\aap] {10.1051/0004-6361/201321397}, \href {https://ui.adsabs.harvard.edu/abs/2013A&A...560A.109H} {560, A109}

\bibitem[\protect\citeauthoryear{{Helmi}, {Babusiaux}, {Koppelman}, {Massari}, {Veljanoski}  \& {Brown}}{{Helmi} et~al.}{2018}]{Helmi+2018}
{Helmi} A.,  {Babusiaux} C.,  {Koppelman} H.~H.,  {Massari} D.,  {Veljanoski} J.,   {Brown} A. G.~A.,  2018, \mn@doi [\nat] {10.1038/s41586-018-0625-x}, \href {https://ui.adsabs.harvard.edu/abs/2018Natur.563...85H} {563, 85}

\bibitem[\protect\citeauthoryear{{Hill} et~al.,}{{Hill} et~al.}{2011}]{Hill+2011}
{Hill} V.,  et~al., 2011, \mn@doi [\aap] {10.1051/0004-6361/200913757}, \href {https://ui.adsabs.harvard.edu/abs/2011A&A...534A..80H} {534, A80}

\bibitem[\protect\citeauthoryear{{Hosek}, {Lu}, {Lam}, {Gautam}, {Lockhart}, {Kim}  \& {Jia}}{{Hosek} et~al.}{2020}]{SPISEA}
{Hosek} Matthew~W. J.,  {Lu} J.~R.,  {Lam} C.~Y.,  {Gautam} A.~K.,  {Lockhart} K.~E.,  {Kim} D.,   {Jia} S.,  2020, \mn@doi [\aj] {10.3847/1538-3881/aba533}, \href {https://ui.adsabs.harvard.edu/abs/2020AJ....160..143H} {160, 143}

\bibitem[\protect\citeauthoryear{{Ibata}, {Chapman}, {Ferguson}, {Irwin}, {Lewis}  \& {McConnachie}}{{Ibata} et~al.}{2004}]{Ibata+2004}
{Ibata} R.,  {Chapman} S.,  {Ferguson} A.~M.~N.,  {Irwin} M.,  {Lewis} G.,   {McConnachie} A.,  2004, \mn@doi [\mnras] {10.1111/j.1365-2966.2004.07759.x}, \href {https://ui.adsabs.harvard.edu/abs/2004MNRAS.351..117I} {351, 117}

\bibitem[\protect\citeauthoryear{{Imig} et~al.,}{{Imig} et~al.}{2023}]{Imig+2023}
{Imig} J.,  et~al., 2023, \mn@doi [\apj] {10.3847/1538-4357/ace9b8}, \href {https://ui.adsabs.harvard.edu/abs/2023ApJ...954..124I} {954, 124}

\bibitem[\protect\citeauthoryear{{Ivezi{\'c}} et~al.,}{{Ivezi{\'c}} et~al.}{2019}]{LSST}
{Ivezi{\'c}} {\v{Z}}.,  et~al., 2019, \mn@doi [\apj] {10.3847/1538-4357/ab042c}, \href {https://ui.adsabs.harvard.edu/abs/2019ApJ...873..111I} {873, 111}

\bibitem[\protect\citeauthoryear{{Kollmeier} et~al.,}{{Kollmeier} et~al.}{2017}]{SDSS-V}
{Kollmeier} J.~A.,  et~al., 2017, arXiv e-prints, \href {https://ui.adsabs.harvard.edu/abs/2017arXiv171103234K} {p. arXiv:1711.03234}

\bibitem[\protect\citeauthoryear{{Kormendy} \& {Bender}}{{Kormendy} \& {Bender}}{1999}]{Kormendy+Bender1999}
{Kormendy} J.,  {Bender} R.,  1999, \mn@doi [\apj] {10.1086/307665}, \href {https://ui.adsabs.harvard.edu/abs/1999ApJ...522..772K} {522, 772}

\bibitem[\protect\citeauthoryear{{Kruijssen}, {Pfeffer}, {Reina-Campos}, {Crain}  \& {Bastian}}{{Kruijssen} et~al.}{2019}]{Kruijssen+2019}
{Kruijssen} J.~M.~D.,  {Pfeffer} J.~L.,  {Reina-Campos} M.,  {Crain} R.~A.,   {Bastian} N.,  2019, \mn@doi [\mnras] {10.1093/mnras/sty1609}, \href {https://ui.adsabs.harvard.edu/abs/2019MNRAS.486.3180K} {486, 3180}

\bibitem[\protect\citeauthoryear{{Lang}}{{Lang}}{2014}]{unWISE1}
{Lang} D.,  2014, \mn@doi [\aj] {10.1088/0004-6256/147/5/108}, \href {https://ui.adsabs.harvard.edu/abs/2014AJ....147..108L} {147, 108}

\bibitem[\protect\citeauthoryear{{Lian} et~al.,}{{Lian} et~al.}{2020}]{Lian+2020a}
{Lian} J.,  et~al., 2020, \mn@doi [\mnras] {10.1093/mnras/staa2078}, \href {https://ui.adsabs.harvard.edu/abs/2020MNRAS.497.2371L} {497, 2371}

\bibitem[\protect\citeauthoryear{{Lian}, {Zasowski}, {Chen}, {Imig}, {Wang}, {Boardman}  \& {Liu}}{{Lian} et~al.}{2024}]{Lian_2024_MWsize}
{Lian} J.,  {Zasowski} G.,  {Chen} B.,  {Imig} J.,  {Wang} T.,  {Boardman} N.,   {Liu} X.,  2024, arXiv e-prints, p. arXiv:2406.05604

\bibitem[\protect\citeauthoryear{{Licquia} \& {Newman}}{{Licquia} \& {Newman}}{2015}]{Licquia+Newman2015}
{Licquia} T.~C.,  {Newman} J.~A.,  2015, \mn@doi [\apj] {10.1088/0004-637X/806/1/96}, \href {https://ui.adsabs.harvard.edu/abs/2015ApJ...806...96L} {806, 96}

\bibitem[\protect\citeauthoryear{{Licquia}, {Newman}  \& {Bershady}}{{Licquia} et~al.}{2016}]{Licquia+2016}
{Licquia} T.~C.,  {Newman} J.~A.,   {Bershady} M.~A.,  2016, \mn@doi [\apj] {10.3847/1538-4357/833/2/220}, \href {https://ui.adsabs.harvard.edu/abs/2016ApJ...833..220L} {833, 220}

\bibitem[\protect\citeauthoryear{{Mackereth} et~al.,}{{Mackereth} et~al.}{2019}]{Mackereth_2019_diskheating}
{Mackereth} J.~T.,  et~al., 2019, \mnras, 489, 176

\bibitem[\protect\citeauthoryear{{Majewski} et~al.,}{{Majewski} et~al.}{2017}]{Majewski+2017}
{Majewski} S.~R.,  et~al., 2017, \mn@doi [\aj] {10.3847/1538-3881/aa784d}, \href {https://ui.adsabs.harvard.edu/abs/2017AJ....154...94M} {154, 94}

\bibitem[\protect\citeauthoryear{{Martig}, {Minchev}, {Ness}, {Fouesneau}  \& {Rix}}{{Martig} et~al.}{2016}]{Martig+2016}
{Martig} M.,  {Minchev} I.,  {Ness} M.,  {Fouesneau} M.,   {Rix} H.-W.,  2016, \mn@doi [\apj] {10.3847/0004-637X/831/2/139}, \href {https://ui.adsabs.harvard.edu/abs/2016ApJ...831..139M} {831, 139}

\bibitem[\protect\citeauthoryear{{McConnachie}, {Irwin}, {Ferguson}, {Ibata}, {Lewis}  \& {Tanvir}}{{McConnachie} et~al.}{2005}]{M31Distance}
{McConnachie} A.~W.,  {Irwin} M.~J.,  {Ferguson} A.~M.~N.,  {Ibata} R.~A.,  {Lewis} G.~F.,   {Tanvir} N.,  2005, \mn@doi [\mnras] {10.1111/j.1365-2966.2004.08514.x}, \href {https://ui.adsabs.harvard.edu/abs/2005MNRAS.356..979M} {356, 979}

\bibitem[\protect\citeauthoryear{{Meisner}, {Lang}  \& {Schlegel}}{{Meisner} et~al.}{2017a}]{unWISE2}
{Meisner} A.~M.,  {Lang} D.,   {Schlegel} D.~J.,  2017a, \mn@doi [\aj] {10.3847/1538-3881/153/1/38}, \href {https://ui.adsabs.harvard.edu/abs/2017AJ....153...38M} {153, 38}

\bibitem[\protect\citeauthoryear{{Meisner}, {Lang}  \& {Schlegel}}{{Meisner} et~al.}{2017b}]{unWISE3}
{Meisner} A.~M.,  {Lang} D.,   {Schlegel} D.~J.,  2017b, \mn@doi [\aj] {10.3847/1538-3881/aa894e}, \href {https://ui.adsabs.harvard.edu/abs/2017AJ....154..161M} {154, 161}

\bibitem[\protect\citeauthoryear{{Myeong}, {Vasiliev}, {Iorio}, {Evans}  \& {Belokurov}}{{Myeong} et~al.}{2019}]{Myeong+2019}
{Myeong} G.~C.,  {Vasiliev} E.,  {Iorio} G.,  {Evans} N.~W.,   {Belokurov} V.,  2019, \mn@doi [\mnras] {10.1093/mnras/stz1770}, \href {https://ui.adsabs.harvard.edu/abs/2019MNRAS.488.1235M} {488, 1235}

\bibitem[\protect\citeauthoryear{{Ness} \& {Lang}}{{Ness} \& {Lang}}{2016}]{Ness+Lang2016}
{Ness} M.,  {Lang} D.,  2016, \mn@doi [\aj] {10.3847/0004-6256/152/1/14}, \href {https://ui.adsabs.harvard.edu/abs/2016AJ....152...14N} {152, 14}

\bibitem[\protect\citeauthoryear{{Ness} et~al.,}{{Ness} et~al.}{2013}]{Ness+2013}
{Ness} M.,  et~al., 2013, \mn@doi [\mnras] {10.1093/mnras/stt533}, \href {https://ui.adsabs.harvard.edu/abs/2013MNRAS.432.2092N} {432, 2092}

\bibitem[\protect\citeauthoryear{{Ness}, {Hogg}, {Rix}, {Ho}  \& {Zasowski}}{{Ness} et~al.}{2015}]{TheCannon1}
{Ness} M.,  {Hogg} D.~W.,  {Rix} H.~W.,  {Ho} A. Y.~Q.,   {Zasowski} G.,  2015, \mn@doi [\apj] {10.1088/0004-637X/808/1/16}, \href {https://ui.adsabs.harvard.edu/abs/2015ApJ...808...16N} {808, 16}

\bibitem[\protect\citeauthoryear{{Ness} et~al.,}{{Ness} et~al.}{2016}]{Ness_2016_apogeekinematics}
{Ness} M.,  et~al., 2016, \apj, 819, 2

\bibitem[\protect\citeauthoryear{{Nidever} et~al.,}{{Nidever} et~al.}{2015}]{Nidever+2015}
{Nidever} D.~L.,  et~al., 2015, \mn@doi [\aj] {10.1088/0004-6256/150/6/173}, \href {https://ui.adsabs.harvard.edu/abs/2015AJ....150..173N} {150, 173}

\bibitem[\protect\citeauthoryear{{Nidever} et~al.,}{{Nidever} et~al.}{2024}]{Nidever+2024}
{Nidever} D.~L.,  et~al., 2024, in {Tabatabaei} F.,  {Barbuy} B.,   {Ting} Y.-S.,  eds,  Proceedings of the International Astronomical Union Vol. 377, Early Disk-Galaxy Formation from JWST to the Milky Way. pp 115--122 (\mn@eprint {arXiv} {2306.04688}), \mn@doi{10.1017/S1743921323002016}

\bibitem[\protect\citeauthoryear{{Olsen}, {Blum}, {Stephens}, {Davidge}, {Massey}, {Strom}  \& {Rigaut}}{{Olsen} et~al.}{2006}]{Olsen+2006}
{Olsen} K. A.~G.,  {Blum} R.~D.,  {Stephens} A.~W.,  {Davidge} T.~J.,  {Massey} P.,  {Strom} S.~E.,   {Rigaut} F.,  2006, \mn@doi [\aj] {10.1086/504900}, \href {https://ui.adsabs.harvard.edu/abs/2006AJ....132..271O} {132, 271}

\bibitem[\protect\citeauthoryear{{Opitsch}, {Fabricius}, {Saglia}, {Bender}, {Bla{\~n}a}  \& {Gerhard}}{{Opitsch} et~al.}{2018}]{Opitsch+2018}
{Opitsch} M.,  {Fabricius} M.~H.,  {Saglia} R.~P.,  {Bender} R.,  {Bla{\~n}a} M.,   {Gerhard} O.,  2018, \mn@doi [\aap] {10.1051/0004-6361/201730597}, \href {https://ui.adsabs.harvard.edu/abs/2018A&A...611A..38O} {611, A38}

\bibitem[\protect\citeauthoryear{{Portail}, {Gerhard}, {Wegg}  \& {Ness}}{{Portail} et~al.}{2017}]{Portail+2017}
{Portail} M.,  {Gerhard} O.,  {Wegg} C.,   {Ness} M.,  2017, \mn@doi [\mnras] {10.1093/mnras/stw2819}, \href {https://ui.adsabs.harvard.edu/abs/2017MNRAS.465.1621P} {465, 1621}

\bibitem[\protect\citeauthoryear{{Queiroz} et~al.,}{{Queiroz} et~al.}{2021}]{Queiroz_2021_MWbar}
{Queiroz} A.~B.~A.,  et~al., 2021, \aap, 656, A156

\bibitem[\protect\citeauthoryear{{Reddy}, {Lambert}  \& {Allende Prieto}}{{Reddy} et~al.}{2006}]{Reddy+2006}
{Reddy} B.~E.,  {Lambert} D.~L.,   {Allende Prieto} C.,  2006, \mn@doi [\mnras] {10.1111/j.1365-2966.2006.10148.x}, \href {https://ui.adsabs.harvard.edu/abs/2006MNRAS.367.1329R} {367, 1329}

\bibitem[\protect\citeauthoryear{{Rojas-Arriagada}, {Zoccali}, {Schultheis}, {Recio-Blanco}, {Zasowski}, {Minniti}, {J{\"o}nsson}  \& {Cohen}}{{Rojas-Arriagada} et~al.}{2019}]{Rojas-Arriagada+2019}
{Rojas-Arriagada} A.,  {Zoccali} M.,  {Schultheis} M.,  {Recio-Blanco} A.,  {Zasowski} G.,  {Minniti} D.,  {J{\"o}nsson} H.,   {Cohen} R.~E.,  2019, \mn@doi [\aap] {10.1051/0004-6361/201834126}, \href {https://ui.adsabs.harvard.edu/abs/2019A&A...626A..16R} {626, A16}

\bibitem[\protect\citeauthoryear{{Rojas-Arriagada} et~al.,}{{Rojas-Arriagada} et~al.}{2020}]{Rojas-Arriagada+2020}
{Rojas-Arriagada} A.,  et~al., 2020, \mn@doi [\mnras] {10.1093/mnras/staa2807}, \href {https://ui.adsabs.harvard.edu/abs/2020MNRAS.499.1037R} {499, 1037}

\bibitem[\protect\citeauthoryear{{Sadoun}, {Mohayaee}  \& {Colin}}{{Sadoun} et~al.}{2014}]{Sadoun+2014}
{Sadoun} R.,  {Mohayaee} R.,   {Colin} J.,  2014, \mn@doi [\mnras] {10.1093/mnras/stu850}, \href {https://ui.adsabs.harvard.edu/abs/2014MNRAS.442..160S} {442, 160}

\bibitem[\protect\citeauthoryear{{Saglia} et~al.,}{{Saglia} et~al.}{2010}]{Saglia+2010}
{Saglia} R.~P.,  et~al., 2010, \mn@doi [\aap] {10.1051/0004-6361/200912805}, \href {https://ui.adsabs.harvard.edu/abs/2010A&A...509A..61S} {509, A61}

\bibitem[\protect\citeauthoryear{{Saglia}, {Opitsch}, {Fabricius}, {Bender}, {Bla{\~n}a}  \& {Gerhard}}{{Saglia} et~al.}{2018}]{Saglia+2018}
{Saglia} R.~P.,  {Opitsch} M.,  {Fabricius} M.~H.,  {Bender} R.,  {Bla{\~n}a} M.,   {Gerhard} O.,  2018, \mn@doi [\aap] {10.1051/0004-6361/201732517}, \href {https://ui.adsabs.harvard.edu/abs/2018A&A...618A.156S} {618, A156}

\bibitem[\protect\citeauthoryear{{Santana} et~al.,}{{Santana} et~al.}{2021}]{Santana+2021}
{Santana} F.~A.,  et~al., 2021, \mn@doi [\aj] {10.3847/1538-3881/ac2cbc}, \href {https://ui.adsabs.harvard.edu/abs/2021AJ....162..303S} {162, 303}

\bibitem[\protect\citeauthoryear{{Sattler}, {Pinna}, {Neumayer}, {Falc{\'o}n-Barroso}, {Martig}, {Gadotti}, {van de Ven}  \& {Minchev}}{{Sattler} et~al.}{2023}]{Sattler+2023}
{Sattler} N.,  {Pinna} F.,  {Neumayer} N.,  {Falc{\'o}n-Barroso} J.,  {Martig} M.,  {Gadotti} D.~A.,  {van de Ven} G.,   {Minchev} I.,  2023, \mn@doi [\mnras] {10.1093/mnras/stad275}, \href {https://ui.adsabs.harvard.edu/abs/2023MNRAS.520.3066S} {520, 3066}

\bibitem[\protect\citeauthoryear{{Saydjari} et~al.,}{{Saydjari} et~al.}{2023}]{DECaPS2}
{Saydjari} A.~K.,  et~al., 2023, \mn@doi [\apjs] {10.3847/1538-4365/aca594}, \href {https://ui.adsabs.harvard.edu/abs/2023ApJS..264...28S} {264, 28}

\bibitem[\protect\citeauthoryear{{Schlafly} et~al.,}{{Schlafly} et~al.}{2018}]{DECaPS}
{Schlafly} E.~F.,  et~al., 2018, \mn@doi [\apjs] {10.3847/1538-4365/aaa3e2}, \href {https://ui.adsabs.harvard.edu/abs/2018ApJS..234...39S} {234, 39}

\bibitem[\protect\citeauthoryear{{Sch{\"o}nrich} \& {Binney}}{{Sch{\"o}nrich} \& {Binney}}{2009}]{Schonrich+Binney2009}
{Sch{\"o}nrich} R.,  {Binney} J.,  2009, \mn@doi [\mnras] {10.1111/j.1365-2966.2009.14750.x}, \href {https://ui.adsabs.harvard.edu/abs/2009MNRAS.396..203S} {396, 203}

\bibitem[\protect\citeauthoryear{{Scott}, {van de Sande}, {Sharma}, {Bland-Hawthorn}, {Freeman}, {Gerhard}, {Hayden}  \& {McDermid}}{{Scott} et~al.}{2021}]{Scott+2021}
{Scott} N.,  {van de Sande} J.,  {Sharma} S.,  {Bland-Hawthorn} J.,  {Freeman} K.,  {Gerhard} O.,  {Hayden} M.~R.,   {McDermid} R.,  2021, \mn@doi [\apjl] {10.3847/2041-8213/abfc57}, \href {https://ui.adsabs.harvard.edu/abs/2021ApJ...913L..11S} {913, L11}

\bibitem[\protect\citeauthoryear{{Semenov}, {Conroy}, {Chandra}, {Hernquist}  \& {Nelson}}{{Semenov} et~al.}{2024}]{Semenov+2024}
{Semenov} V.~A.,  {Conroy} C.,  {Chandra} V.,  {Hernquist} L.,   {Nelson} D.,  2024, \mn@doi [\apj] {10.3847/1538-4357/ad150a}, \href {https://ui.adsabs.harvard.edu/abs/2024ApJ...962...84S} {962, 84}

\bibitem[\protect\citeauthoryear{{Sharma} et~al.,}{{Sharma} et~al.}{2021a}]{Sharma+2021a}
{Sharma} S.,  et~al., 2021a, \mn@doi [\mnras] {10.1093/mnras/stab1086}, \href {https://ui.adsabs.harvard.edu/abs/2021MNRAS.506.1761S} {506, 1761}

\bibitem[\protect\citeauthoryear{{Sharma}, {Hayden}  \& {Bland-Hawthorn}}{{Sharma} et~al.}{2021b}]{Sharma+2021b}
{Sharma} S.,  {Hayden} M.~R.,   {Bland-Hawthorn} J.,  2021b, \mn@doi [\mnras] {10.1093/mnras/stab2015}, \href {https://ui.adsabs.harvard.edu/abs/2021MNRAS.507.5882S} {507, 5882}

\bibitem[\protect\citeauthoryear{{Skrutskie} et~al.,}{{Skrutskie} et~al.}{2006}]{2MASS}
{Skrutskie} M.~F.,  et~al., 2006, \mn@doi [\aj] {10.1086/498708}, \href {https://ui.adsabs.harvard.edu/abs/2006AJ....131.1163S} {131, 1163}

\bibitem[\protect\citeauthoryear{{Spitoni}, {Silva Aguirre}, {Matteucci}, {Calura}  \& {Grisoni}}{{Spitoni} et~al.}{2019}]{Spitoni+2019}
{Spitoni} E.,  {Silva Aguirre} V.,  {Matteucci} F.,  {Calura} F.,   {Grisoni} V.,  2019, \mn@doi [\aap] {10.1051/0004-6361/201834188}, \href {https://ui.adsabs.harvard.edu/abs/2019A&A...623A..60S} {623, A60}

\bibitem[\protect\citeauthoryear{{Stewart}, {Bullock}, {Wechsler}, {Maller}  \& {Zentner}}{{Stewart} et~al.}{2008}]{Stewart+2008}
{Stewart} K.~R.,  {Bullock} J.~S.,  {Wechsler} R.~H.,  {Maller} A.~H.,   {Zentner} A.~R.,  2008, \mn@doi [\apj] {10.1086/588579}, \href {https://ui.adsabs.harvard.edu/abs/2008ApJ...683..597S} {683, 597}

\bibitem[\protect\citeauthoryear{Tukey}{Tukey}{1958}]{jackknife}
Tukey J.~W.,  1958, \mn@doi [] {https://doi.org/10.1214/aoms/1177706647}, 29, 614

\bibitem[\protect\citeauthoryear{{Vieira}, {Carraro}, {Korchagin}, {Lutsenko}, {Girard}  \& {van Altena}}{{Vieira} et~al.}{2022}]{Vieira_2022_MWdisks}
{Vieira} K.,  {Carraro} G.,  {Korchagin} V.,  {Lutsenko} A.,  {Girard} T.~M.,   {van Altena} W.,  2022, \apj, 932, 28

\bibitem[\protect\citeauthoryear{{Wang}, {Hayden}, {Sharma}, {van de Sande}, {Bland-Hawthorn}, {Vaughan}, {Martig}  \& {Pinna}}{{Wang} et~al.}{2023}]{Wang+2023}
{Wang} Z.,  {Hayden} M.~R.,  {Sharma} S.,  {van de Sande} J.,  {Bland-Hawthorn} J.,  {Vaughan} S.,  {Martig} M.,   {Pinna} F.,  2023, \mn@doi [arXiv e-prints] {10.48550/arXiv.2310.18258}, \href {https://ui.adsabs.harvard.edu/abs/2023arXiv231018258W} {p. arXiv:2310.18258}

\bibitem[\protect\citeauthoryear{{Weinberg} et~al.,}{{Weinberg} et~al.}{2019}]{Weinberg+2019}
{Weinberg} D.~H.,  et~al., 2019, \mn@doi [\apj] {10.3847/1538-4357/ab07c7}, \href {https://ui.adsabs.harvard.edu/abs/2019ApJ...874..102W} {874, 102}

\bibitem[\protect\citeauthoryear{{Wilson} et~al.,}{{Wilson} et~al.}{2019}]{Wilson+2019}
{Wilson} J.~C.,  et~al., 2019, \mn@doi [\pasp] {10.1088/1538-3873/ab0075}, \href {https://ui.adsabs.harvard.edu/abs/2019PASP..131e5001W} {131, 055001}

\bibitem[\protect\citeauthoryear{{Wu}, {Xiang}, {Chen}, {Zhao}, {Bi}, {Li}, {Li}  \& {Huang}}{{Wu} et~al.}{2021}]{Wu+2021}
{Wu} Y.,  {Xiang} M.,  {Chen} Y.,  {Zhao} G.,  {Bi} S.,  {Li} C.,  {Li} Y.,   {Huang} Y.,  2021, \mn@doi [\mnras] {10.1093/mnras/staa3949}, \href {https://ui.adsabs.harvard.edu/abs/2021MNRAS.501.4917W} {501, 4917}

\bibitem[\protect\citeauthoryear{{Yoshii}}{{Yoshii}}{1982}]{Yoshii1982}
{Yoshii} Y.,  1982, \pasj, \href {https://ui.adsabs.harvard.edu/abs/1982PASJ...34..365Y} {34, 365}

\bibitem[\protect\citeauthoryear{{Zasowski} et~al.,}{{Zasowski} et~al.}{2013}]{Zasowski+2013}
{Zasowski} G.,  et~al., 2013, \mn@doi [\aj] {10.1088/0004-6256/146/4/81}, \href {https://ui.adsabs.harvard.edu/abs/2013AJ....146...81Z} {146, 81}

\bibitem[\protect\citeauthoryear{{Zasowski}, {Ness}, {Garc{\'\i}a P{\'e}rez}, {Martinez-Valpuesta}, {Johnson}  \& {Majewski}}{{Zasowski} et~al.}{2016}]{Zasowski+2016}
{Zasowski} G.,  {Ness} M.~K.,  {Garc{\'\i}a P{\'e}rez} A.~E.,  {Martinez-Valpuesta} I.,  {Johnson} J.~A.,   {Majewski} S.~R.,  2016, \mn@doi [\apj] {10.3847/0004-637X/832/2/132}, \href {https://ui.adsabs.harvard.edu/abs/2016ApJ...832..132Z} {832, 132}

\bibitem[\protect\citeauthoryear{{Zasowski} et~al.,}{{Zasowski} et~al.}{2017}]{Zasowski+2017}
{Zasowski} G.,  et~al., 2017, \mn@doi [\aj] {10.3847/1538-3881/aa8df9}, \href {https://ui.adsabs.harvard.edu/abs/2017AJ....154..198Z} {154, 198}

\bibitem[\protect\citeauthoryear{{Zoccali}, {Hill}, {Lecureur}, {Barbuy}, {Renzini}, {Minniti}, {G{\'o}mez}  \& {Ortolani}}{{Zoccali} et~al.}{2008}]{Zoccali+2008}
{Zoccali} M.,  {Hill} V.,  {Lecureur} A.,  {Barbuy} B.,  {Renzini} A.,  {Minniti} D.,  {G{\'o}mez} A.,   {Ortolani} S.,  2008, \mn@doi [\aap] {10.1051/0004-6361:200809394}, \href {https://ui.adsabs.harvard.edu/abs/2008A&A...486..177Z} {486, 177}

\bibitem[\protect\citeauthoryear{{de Vaucouleurs}, {de Vaucouleurs}, {Corwin}, {Buta}, {Paturel}  \& {Fouque}}{{de Vaucouleurs} et~al.}{1991}]{M31velocity}
{de Vaucouleurs} G.,  {de Vaucouleurs} A.,  {Corwin} Herold~G. J.,  {Buta} R.~J.,  {Paturel} G.,   {Fouque} P.,  1991, {Third Reference Catalogue of Bright Galaxies}

\bibitem[\protect\citeauthoryear{{van de Sande}, {Fraser-McKelvie}, {Fisher}, {Martig}, {Hayden}  \& {Geckos Survey Collaboration}}{{van de Sande} et~al.}{2024}]{GECKOS}
{van de Sande} J.,  {Fraser-McKelvie} A.,  {Fisher} D.~B.,  {Martig} M.,  {Hayden} M.~R.,   {Geckos Survey Collaboration} 2024, in {Tabatabaei} F.,  {Barbuy} B.,   {Ting} Y.-S.,  eds,  Proceedings of the International Astronomical Union Vol. 377, Early Disk-Galaxy Formation from JWST to the Milky Way. pp 27--33 (\mn@eprint {arXiv} {2306.00059}), \mn@doi{10.1017/S1743921323001138}

\makeatother
\end{thebibliography}






\bsp	
\label{lastpage}
\end{document}